\documentstyle[12pt,citesort]{article}
\catcode`\@=11

\newcount\hour
\newcount\minute
\newtoks\amorpm \hour=\time\divide\hour by 60\minute
=\time{\multiply\hour by 60 \global\advance\minute by-\hour}
\edef\standardtime{{\ifnum\hour<12 \global\amorpm={am}%
        \else\global\amorpm={pm}\advance\hour by-12 \fi
        \ifnum\hour=0 \hour=12 \fi
        \number\hour:\ifnum\minute<10
        0\fi\number\minute\the\amorpm}}
\edef\militarytime{\number\hour:\ifnum\minute<10
0\fi\number\minute}

\def\draftlabel#1{{\@bsphack\if@filesw {\let\thepage\relax
   \xdef\@gtempa{\write\@auxout{\string
      \newlabel{#1}{{\@currentlabel}{\thepage}}}}}\@gtempa
   \if@nobreak \ifvmode\nobreak\fi\fi\fi\@esphack}
        \gdef\@eqnlabel{#1}}
\def\@eqnlabel{}
\def\@vacuum{}
\def\marginnote#1{}
\def\draftmarginnote#1{\marginpar{\raggedright\scriptsize\tt#1}}
 \def \lc {light-cone\ }

\def\draft{
        \pagestyle{plain}
        \overfullrule=2pt
        \oddsidemargin -.5truein
        \def\@oddhead{\sl \phantom{\today\quad\militarytime} \hfil
        \smash{\Large\sl DRAFT} \hfil \today\quad\militarytime}
        \let\@evenhead\@oddhead
        \let\label=\draftlabel
        \let\marginnote=\draftmarginnote
        \def\ps@empty{\let\@mkboth\@gobbletwo
        \def\@oddfoot{\hfil \smash{\Large\sl DRAFT} \hfil}
        \let\@evenfoot\@oddhead}
        \def\@eqnnum{(\theequation)\rlap{\kern\marginparsep\tt\@eqnlabel}%
        \global\let\@eqnlabel\@vacuum}  }

\newcommand{\rf}[1]{(\ref{#1})}
\renewcommand{\theequation}{\thesection.\arabic{equation}}
\renewcommand{\thefootnote}{\fnsymbol{footnote}}

\newcommand{\newsection}{    % Numeration of eqs. is automatic
\setcounter{equation}{0}
\section}
\def\appendix#1{
  \addtocounter{section}{1}
  \setcounter{equation}{0}
  \renewcommand{\thesection}{\Alph{section}}
  \section*{Appendix \thesection\protect\indent \parbox[t]{11.15cm}
  {#1} }
  \addcontentsline{toc}{section}{Appendix \thesection\ \ \ #1}
  }

\def\apr{{{\rm A}^\prime}}
\def\bpr{{{\rm B}^\prime}}
\def\cpr{{{\rm C}^\prime}}

\def\sca{{\scriptscriptstyle{\cal  A}}}
\def\scb{{\scriptscriptstyle{\cal  B}}}

\def\Csp{C^\prime}

\def\al{{\lambda}}

 \def \const {{\rm const}}

\def\vm{{\mu}}
\def\vn{{\nu}}

\def\sfa{{\sf a}}

\def\dsfa{\dot{\sf a}}
\def\dsfb{\dot{\sf b}}

\def \td {\tilde }

\def \gg {{\cal g}}

\def \foot {\footnote}
\def \bi{\bibitem}
\def \la {\label}

\def \ha {{1 \over 2}}

\def \ov {\over}

\def\nline{\,\nabla\kern -0.7em\raise0.2ex\hbox{/}\,\,}
\def\yline{\,y\kern -0.47em /}
\def\aline{\,a\kern -0.49em /}
\def\parline{\,\partial\kern -0.55em /\,\,}

\def \t {\theta}
\def \s{\sigma}

\def\NPB#1(#2)#3{{\it Nucl. Phys.} {\bf B#1} (#2) #3}
\def\PRD#1(#2)#3{{\it Phys. Rev.} {\bf D#1} (#2) #3}
\def\PLB#1(#2)#3{{\it Phys. Lett.} {\bf B#1} (#2) #3}

\def\RMP#1(#2)#3{{\it Rev. Mod. Phys.} {\bf #1} (#2) #3}
\def\MPLA#1(#2)#3{{\it Mod. Phys. Lett.} {\bf A#1} (#2) #3}
\def\CQG#1(#2)#3{{\it Class. Quantum Grav.} {\bf #1} (#2) #3}
\def\AP#1(#2)#3{{\it Ann. Phys.} {\bf #1} (#2) #3}
\def\SJNP#1(#2)#3{{\it Sov. J. Nucl. Phys.} {\bf #1} (#2) #3}

\def \del{\partial}

\def\det{\hbox{det}}
\def\be{\begin{equation}}
\def\ee{\end{equation}}

\def \ci {\cite}
\def \g {\gamma}

\def \k {\kappa}
\def \l {\lambda}

\def\apr{{A'}}
\def \m {\mu}
\def \n {\nu}

\def\x'{\mathaccent 19 x}
\def\y'{\mathaccent 19 y}
\def\n'{\mathaccent 19 n}
\def\u'{\mathaccent 19 u}

\def\X'{\mathaccent 19 X}
\def\Y'{\mathaccent 19 Y}
\def\Z'{\mathaccent 19 Z}

\def\et'{\mathaccent 19 \eta}
\def\th'{\mathaccent 19 \theta}
\def\lam'{\mathaccent 19 \lambda}
\def\varet'{\mathaccent 19 \vartheta}
\def\rh'{\mathaccent 19 \rho}
\def\ph'{\mathaccent 19 \phi}
\def\xb'{\mathaccent 19 {\bar{x}}}

\begin{document}
%\draft
 %%%%%%%%%%%%%%%%%%%%%%%%%%%%%%%%%%%%%%%%%%%%%%%%%%%

\begin{titlepage}
\begin{flushright}
FIAN/TD/00-14
\\
OHSTPY-HEP-T-00-017
\\
UFIFT-HEP-00-26\\
hep-th/0009171\\
\end{flushright}
\vspace{.5cm}

\begin{center}
{\LARGE
Light-cone Superstring in  AdS Space-time
%$\k$-symmetry light cone  gauge
%Superstring action  in AdS$_5 \times $S$^5$:\\[.1cm]
  %       $\k$-symmetry    light cone  gauge 
   }\\[.2cm]
\vspace{1.1cm}
{\large R.R. Metsaev,${}^{{\rm a,b,}}$\footnote{\
E-mail: metsaev@lpi.ru, metsaev@pacific.mps.ohio-state.edu}
C. B. Thorn${}^{{\rm c,}}$\footnote{\ E-mail  address: 
thorn@phys.ufl.edu}
and A.A. Tseytlin${}^{{\rm a,}}$\footnote{\ Also at 
Imperial College, London and   Lebedev Institute, Moscow.\
 E-mail: tseytlin.1@osu.edu} }

\vspace{18pt}
 ${}^{{\rm a\ }}${\it
 Department of Physics,
The Ohio State University  \\
Columbus, OH 43210-1106, USA\\
}

\vspace{6pt}

${}^{{\rm b\ }}${\it
Department of Theoretical Physics, P.N. Lebedev Physical
Institute,\\ Leninsky prospect 53,  Moscow 117924, Russia
}

\vspace{6pt}

${}^{{\rm c\ }}${\it Institute for Fundamental Theory\\
Department of Physics, University of Florida,
Gainesville, FL 32611, USA
}

\end{center}

\vspace{2cm}

\begin{abstract}
%{\bf Abstract} \end{center}
We consider fixing the  bosonic light-cone gauge 
for string in AdS space in the 
%%% phase space  framework, i.e. $x^+ = \tau, \ P^+= $const.
phase space  framework, i.e. by choosing $x^+ = \tau$,
and by choosing $\sigma$ so that $P^+$ is distributed
uniformly (its density ${\cal P}^+$ is independent of 
$\sigma$).
%%%
We discuss classical 
bosonic string in $AdS_d$  and superstring  in 
$AdS_5 \times S^5$. In the latter case the starting point 
is the action found in  hep-th/0007036  where  the 
$\kappa$-symmetry is fixed by a fermionic light-cone type  gauge.
We derive the light-cone Hamiltonian in the 
 $AdS_5 \times S^5$ case and in the case of superstring in 
 $AdS_3 \times S^3$.
We also obtain a realization of the generators  of the 
 basic superalgebra 
$psu(2,2|4)$ in terms of  $AdS_5 \times S^5$ superstring 
coordinate 2-d fields in the light-cone gauge.
\end{abstract}

\end{titlepage}
\setcounter{page}{1}
\renewcommand{\thefootnote}{\arabic{footnote}}
\setcounter{footnote}{0}

\def \adss {$AdS_5 \times S^5$\ }
\def \N {{\cal N}}
\def \lc {light-cone\ }
\def \ta { \tau}
\def \s { \sigma }
\def  \gg  { {\rm g}}
\def \sg {\sqrt {g }}
\def \te {\theta}
\def \vp {\varphi}
\def \gij {g_{ab}}
\def \xp {x^+}
\def \xm {x^-}
\def \p {\phi}
\def \vt {\theta}
\def \bx {\bar x} \def \a { \alpha}
\def \r {\rho}
\def \fourth {{1 \ov 4}}
\def \DD {{\cal D}}
\def \half {{1 \ov 2}}
\def \inv {^{-1}}
\def \ri {{\rm i}}
\def \D {{\cal D}}
\def \DD {{\rm D}}
\def \vr {\varrho}
\def \diag {{\rm diag}} \def \td { \tilde }
\def \tta {\td \eta}
\def \cA {{\cal A}}
\def \cB   {{\cal B}}
\def \na {\nabla}
\def \al {\perp}
\def \PP {{\cal P}} 

%%%%%%%%%%%%%%%%%%%%%%%%%%%%%%%%%%%%%%%%%%%
\newsection{Introduction}
%%%%%%%%%%%%%%%%%%%%%%%%%%%%%%

%To hope to understand string theory duals of 
To better understand string theory duals of
%%% 
gauge theories with various amounts of supersymmetry
\ci{Pol}
it is important to make progress in what should  be the 
simplest case --  (large $N$) $\cal N$=4 Super Yang-Mills 
 theory dual 
to (weakly coupled) type IIB superstring theory in \adss space 
with Ramond-Ramond flux \ci{mald}.
%Though this background has a lot of symmetry, to solve
Though this background has a lot of symmetry, solving
%%%%
the corresponding string theory appears   to be a 
%complicated problem. The usual procedure in the known 
complicated problem. The commonly used procedure, in the known 
exactly solvable cases,
 is to start with
 a  string  action, solve the classical string equations, 
 then quantize the theory, find the string spectrum, 
 vertex operators, scattering amplitudes, etc.
Even the first steps in this program are nontrivial in the 
\adss string case.

To start with, the classical
equations of  bosonic string in $AdS$
 space, while completely  integrable, are not explicitly solvable
 (in contrast to, e.g.,  the case of  
 string on a group manifold described by the WZW
 model). 
%The presence of a curved R-R background implies that one should use 
The presence of a curved R-R background indicates that one should use 
the manifestly supersymmetric Green-Schwarz \ci{GS} description 
%of the superstring. Finding an explicit expression
of superstring. Finding an explicit expression
for the curved space GS  action  is 
%known to be hard
difficult
in general
 (one needs to know the component
expansion  of the  background supergravity
superfields). In the present \adss case, this technical problem
 has a nice geometrical
solution based on viewing string as moving on  the supercoset
$PSU(2,2|4)/[SO(1,4) \times SO(5)]$  which replaces the flat superspace
(10-d super Poincare)/$SO(1,9)$  \ci{MT1}. The resulting 
 action, though  explicitly  known \ci{MT1,KRR,MT2}, is highly nonlinear
 containing  terms of many  orders in $\t$. 
 The fermionic part of the action simplifies dramatically 
 in  proper $\k$-symmetry gauges -- it   becomes  quadratic
 and quartic in $\t$ only \ci{Pes,KR,KT}. Its 
 structure is similar to that of the flat space   GS action 
 in a covariant $\k$-symmetry gauge  which  is also 
 quartic in fermions (see e.g.\cite{kav}).

 %%%%%%%%%%%%%%%%%%%%%%%%%%%%%
\subsection{Review}
%%%%%%%%%%%%%%%%%%%%%%

  To illustrate this point, let us consider
 the 4-d Lorentz covariant ``S-gauge" \ci{MT3} which 
 leads to an action equivalent to the one in 
 \ci{Pes,KR} but  
 uses  the fermionic parametrization 
adopted in \ci{MT3}  and here, which is 
 useful for comparison with the \lc gauge 
  actions below
  (see Appendix C of \ci{MT3} for details).
 If one interprets the   \adss supergroup  $PSU(2,2|4)$ as 
 the   $\N=4$ 
 superconformal group in 4 dimensions, it is natural to split 
 the  fermionic generators into 4 standard supergenerators $Q_i$ and 
 4 special conformal supergenerators $S_i$ (we suppress the  
 4-d spinor indices). The associated 
 superstring coordinates will be denoted as 
 $\theta_i$ and $\eta_i$. The covariant ``S-gauge" 
 corresponds to setting all $\eta_i$ to zero. The resulting 
 superstring Lagrangian  written in the ``4+6" parametrization 
 in which the metric of \adss is 
 $ds^2 = Y^2 dx^a dx^a + Y^{-2} dY^MdY^M$ ($a=0,...,3; \ M=1,...,6$)
 has the following simple structure \ci{MT3}\foot{The actions in 
 \ci{Pes,KR} have an isomorphic form, corresponding to a specific choice
 of the 10-d Dirac matrix representation.}
 \begin{eqnarray}
 L &=& - \frac{1}{2}\sqrt g g^{\vm\vn}\Bigl( Y^2 L_\vm^a L_\vn^a
 + Y^{-2}\del_\vm Y^M \del_\vn Y^M \Bigr)
  - \bigl( {\rm i} \epsilon^{\mu\nu} \del_\mu Y^M 
   \theta^{\dsfa i} \rho^M_{ij} \del_\nu  \theta_{\dsfa}^j +
   h.c.\bigr)  \  , 
 \la{bss}
   \\
   &&
 L_\vm^a \equiv \del_\vm x^a
  - \bigl({\rm i} \theta_{\sfa i} 
  (\sigma^a)^{\sfa\dsfb}\del_\vm \theta_{\dsfb}^i 
     + h.c.\bigr) \  , 
     \nonumber
 \end{eqnarray}
 where $\sfa,\dsfb$ are   the  $sl(2,C)$ 
 (4-d spinor) indices and 
 $\theta_{\sfa i}^\dagger = -\theta_{\dsfa}^i$, 
 $\theta^{\sfa}_i{}^\dagger = \theta^{\dsfa i}$. 
 The $\sigma^a$ are 4 
 Pauli matrices (off-diagonal blocks
 of 4-d Dirac matrices in Weyl representation), 
 and the $\rho^M_{ij}$ are similar off-diagonal blocks
 of $SO(6)$ Dirac matrices in chiral representation (see Appendix A).
%%NEW%%
As in  most of the discussion  below  we set the radius $R$ of 
$AdS_5$ and $S^5$   to 1. 

This  covariant $\k$-symmetry gauge fixed  action is 
 well-defined and useful for developing perturbation 
theory near classical ``long'' string  configurations 
ending at the boundary of $AdS_5$ \ci{KT,Forste,DGT}, e.g., 
 the ones appearing in the Wilson loop computations
\ci{malda}. However, the  kinetic term of the fermions 
which has the structure 
$  \del x \theta  \del \theta $
 is degenerate for ``short'' strings \ci{KT}, and thus 
 this action is not directly applicable 
  for  computing  the spectrum of closed string in the bulk 
  of \adss.
  
In order to  avoid  this degeneracy problem,
 it is natural to try to follow 
the same approach which worked remarkably well 
 in flat space \ci{GS,GSlc}:  (i) find a light-cone type 
  $\k$-symmetry gauge in  which the fermion kinetic term 
  becomes $  \del x^+ \theta  \del \theta $, i.e. it involves only one 
  combination $x^+$ of 4-d coordinates, and  then 
   (ii)   choose the  light-cone 
  bosonic gauge $x^+=\tau$. 
  
In the previous paper \ci{MT3} it was shown  how 
the  light-cone type $\k$-symmetry gauge can be fixed
in the original action of \ci{MT1} so that 
 the resulting action   has indeed the required form.\foot{For 
 a different approach to  light-cone type 
  $\k$-symmetry gauge fixing in the \adss string action see 
   \ci{pes}.}
In contrast to the Lorentz covariant ``S-gauge" where 16 fermions 
 $\eta_i$ are set equal to zero, the light-cone 
 gauge  used in  \ci{MT3} corresponds to setting to zero 
 ``half" of the 16 $\theta_i$ and ``half" of the 16 $\eta_i$
 (``half" is defined with respect to $SO(1,1)$ rotations in 
 light-cone directions). The remaining fermions
 are  again denoted  by  $\theta_i, \eta_i$,
 but now they are simply 4+4  complex anticommuting variables 
 carrying no extra  Lorentz indices.
 For comparison,  the flat space  GS action in the 
 light-cone gauge ($ \Gamma^+ \vartheta=0$) 
  \ci{GSlc}, written in a  similar parametrization 
 of the 16 fermionic
 coordinates, has the following form \ci{MT3}
  \be
{\cal L}
=- \frac{1}{2}\sqrt{g}
\partial^\vm x^{\hat{A}}\partial_\vm x^{\hat{A}} - \Bigl[
 \frac{{\rm i}}{2}\sqrt{g} \partial^\vm x^+(
\theta_i\partial_\vm \theta^i
+\eta_i\partial_\vm\eta^i)  - \epsilon^{\vm\vn}
\partial_\vm x^+ \eta^i C_{ij}^\prime \partial_\vn\theta^j+h.c.
\Bigl] \ , \la{sss}
\ee
where $x^{\hat{A}}$ are 10 flat bosonic coordinates
and $C_{ij}^\prime$ is a constant  charge conjugation
matrix. Note that  while the $\theta$'s and $\eta$'s
enter diagonally in the kinetic term,
 they are mixed in the WZ term.
This form of the  original  GS action is the 
flat space limit
 of  the  \lc  \adss action  of \ci{MT3} (cf. \rf{bss})
\begin{eqnarray}
{\cal L} 
&= &
-\sqrt{g}g^{\vm\vn}\Bigl[
Y^2(\partial_\vm x^+ \partial_\vn x ^-
+ \partial_\vm x\partial_\vn\bar{x})
+\frac{1}{2} Y^{-2} D_\mu Y^M D_\nu Y^M  \Bigl]
\nonumber\\
&-&\frac{{\rm i}}{2} \sqrt{g}g^{\vm\vn}
Y^2\partial_\vm x^+
\Bigl[\theta^i\partial_\vn \theta_i
+\theta_i\partial_\vn \theta^i
+\eta^i\partial_\vn \eta_i
+\eta_i\partial_\vn \eta^i 
+{\rm i}  Y^2\partial_\vn x^+(\eta^2)^2\Bigr] 
\nonumber\\
&+&\Bigl[ \epsilon^{\vm\vn}
|Y|\partial_\vm x^+ \eta^i \rho_{ij}^M Y^M
(\partial_\vn\theta^j-{\rm i}\sqrt{2}|Y| \eta^j
\partial_\vn x)+h.c.\Bigl] \ , \label{actki}
\end{eqnarray}
$$ D_\mu  Y^M \equiv  \del_\mu Y^M +  
{\rm i}\eta_i (\rho^{MN})^i{}_j\eta^j  Y^N Y^2 \del_\mu x^+ \ . $$
Here we split the 4-d coordinates as $x^a=(x^+,x^-,x, \bar x)$,
and the matrix $\rho^M$ is the same as in \rf{bss} 
($\rho^{MN}$  is the commutator of $\rho^M$ and its conjugate, 
see Appendix A).

This action  contains  $x^-$  
only in the bosonic part and only linearly, 
and the fermionic kinetic terms are multiplied by 
the derivative of $x^+$ 
only. One expects, therefore, that
after  one fixes  the bosonic \lc gauge, 
 it should lead to 
  a well defined
starting point for quantizing the theory in the 
``short'' string sector.
As was pointed out  in \ci{rudd,pol,MT3} (see also  Section 2), 
in the case of the  $AdS$ type curved spaces, 
the bosonic \lc gauge $x^+=\tau$  in the  Polyakov string action 
can not be combined with the standard conformal gauge
$\sqrt g g^{\m\nu}= \eta^{\mu\nu}$: one needs to impose a condition
on $g_{\m\nu}$ that breaks the 2-d Lorentz symmetry
and leads to a rather  non-standard  string action,
with all terms coupled to the radial function $\phi = \log Y$
of $AdS_5$ space.

%%%%%%%%%%%%%%%%%%%%%%%%%%%%%
\subsection{Summary}
%%%%%%%%%%%%%%%%%%%%%%%%%%%%%%

The  absence  of manifest 2-d Lorentz symmetry  suggests that 
it is natural to use the {\it phase space}  formulation   of 
  the  \lc gauge fixed  theory. To develop such a formulation is the aim 
of  the present paper.
We shall  explain how the original phase space 
GGRT approach 
\ci{ggrt} to \lc gauge fixing  ($x^+=\tau, \ {\cal P}^+=1$) can be 
directly  applied  to  the present case of  a  curved background.

Most of our discussion will be rather formal,
with possible applications depending on insights
into  the
structure of the  resulting  \adss string \lc Hamiltonian. 
To summarize, the \lc phase space Lagrangian, corresponding  to
\rf{actki} that we obtain below,  is given by  (see \rf{phase4}) 
\be
{\cal L} 
= \PP_\al\dot{x}_\al 
+ \PP_M \dot{Y}^M
+\frac{{\rm i}}{2}p^+(\theta^i \dot{\theta}_i
+\eta^i\dot{\eta}_i+\theta_i \dot{\theta}^i+\eta_i\dot{\eta}^i)  + \PP^- \  , 
\la{laah}
\ee
where $\PP_\al, \PP_M$ are the canonical momenta for $x_\al=(x,\bar x)$ 
and $Y^M$,  and 
 the \lc Hamiltonian density  is
$$
{\cal H} = -\PP^- =  
\frac{1}{2p^+}\Bigl[\PP_\al^2 +|Y|^4\PP_M\PP_M
+|Y|^4\x'_\al^2+ \Y'^M\Y'^M
+Y^2(p^{+2}(\eta^2)^2 + 2{\rm i}p^+\eta \rho^{MN}\eta  Y_M \PP_N)\Bigr]
$$
\be
-\ \Bigl[\ |Y|\eta^i \rho_{ij}^M Y^M
(\th'^j - {\rm i}\sqrt{2}|Y| \eta^j\x')+h.c.\Bigr] \  .
\la{haha}
\ee
%This Hamiltonian  should be supplemented by the constraint
As usual for light-cone string, the coordinate $x^-$ does
not appear in the Hamiltonian, but is determined in terms
of the coordinates that do appear via the reparametrization
constraint

\be
p^+\x'^-  + \PP_\al\x'_\al 
+ \PP_M \Y'^M
+\frac{{\rm i}}{2}p^+(
\theta^i\th'_i+\eta^i\et'_i
+\theta_i\th'^i+\eta_i\et'^i) =0   \ .  \la{coons}
\ee
%%%%%
For a closed piece of string, the integral of this constraint over
$\sigma$ constrains the state space to the subspace
invariant under $\sigma$ translations. 
%%%%%

Restoring the dependence on $\a'$ and the  scale  $R$ of \adss
one can then analyze various limits, e.g., 
 (i)  particle theory  limit
($\a' \to 0$),\  (ii) null string limit ($\a' \to \infty$),
 %%NEW%%
\foot{As in the flat space  case 
  \ci{lin},  the null string limit
 is obtained by dropping  terms with derivatives with respect 
 to the world sheet coordinate $\sigma$.}
 (iii) \adss supergravity  or  strong
`t Hooft coupling SYM limit (${\a'/ R^2}\to 0$), \ 
(vi) weak `t Hooft coupling SYM limit  (${\a'/ R^2}\to \infty$). 
%%NEW%%
In the particle theory limit, the string Hamiltonian 
\rf{haha} reduces to the light-cone Hamiltonian for a 
superparticle in \adss found in \ci{met3}. This {\it implies} 
that the ``massless" (zero-mode)  spectrum of the superstring
coincides indeed with the spectrum of type IIB supergravity
compactified on $S^5$!

Further progress depends  on the possibility 
of  making a transformation to some  new variables 
which  may allow one to solve  for the string theory spectrum.
It would also be  interesting 
to make connections to  other related ideas and 
approaches in the literature, 
such as the one in   \ci{ber}.
For example, introducing  twistor-like  variables
in the  present \adss context may turn out to be  useful
(see \ci{kaltw,sor} for some previous discussions of twistors
 in $AdS$ space).\foot{Twistors are very helpful
in  the construction of  
 the theory of  interacting massless higher spin fields
in $AdS$ space
 \ci{vas1,vas4}, which has certain  similarities 
with string theory.}
Another potentially 
promising direction  is  to  
apply  the  methods of integrable systems. It was
demonstrated in \ci{lun,BN1,BN2}  that the  set of the classical 
equations  of  bosonic string in $AdS$
geometry can be interpreted  as a completely 
 integrable system (for related discussions see also 
\ci{veg,frmet}). The integrability property should be true also
for the full system of the classical equations of  
\adss superstring. It should be  crucial to include the  fermions
from the beginning 
  since their coupling to the $AdS_5$ and $S^5$ sets of 
bosons
via the  R-R interaction  terms   insures
the conformal  invariance of the 2-d  string  
theory at the quantum level
 \ci{MT1}. Finally, knowing the light-cone description
of string in $AdS$ space-time provides string theory
guidance for identifying the dynamics that should emerge from
the ``brute force'' approach to a 
field theory/string duality based on directly summing the
planar diagrams of 't Hooft's large $N_c$ limit
\cite{thooftlargen}.
This program is most definitively carried out using a light-cone
parametrization \cite{thornfishnet,browergt,beringrt}, so
its comparison to the results of the present paper 
should be particularly instructive.
For example, there should be a more or less direct link
to the fishnet diagrams \cite{nielsenfishnet} in the
strong 't Hooft coupling limit \cite{thornfishnet,beringrt}.  

%%%%%%%%%%%%%%%%%%%%%%%%%%%%%%%%%%%%%%%%%%%%
\bigskip

The rest of the paper  is organized as follows.
We  start with  the example of the classical bosonic string 
in $AdS$-type space in Section 2. 
%%%NEW%%
We illustrate our phase-space
approach to fixing the light-cone gauge  by deriving the 
 \lc
Hamiltonian for the bosonic string in $AdS_3$ space
with NS-NS flux  described by $SL(2,R)$ WZW model.
In Section 3 we show how some known properties of the
classical  ``long''  bosonic string solutions can be 
directly
understood in the  \lc Hamiltonian  framework of Section 2.
In Section 4 we  review the \lc Hamiltonian 
 description of a
superparticle  in \adss developed in \ci{met3}, which is the zero
slope  limit of the string theory case.
In Section 5  we derive  the 
phase space analog of the \adss superstring Lagrangian 
of \ci{MT3} 
 and the corresponding \lc gauge  Hamiltonian.
There we  use the   form of the action
based on  the ``5+5" parametrization of $AdS_5 \times S^5$, 
while the \lc phase space  counterpart \rf{laah},\rf{haha}
of the  action 
 \rf{actki}
is given in Appendix C.
As an aside, 
we  also  present the \lc   Hamiltonian
for superstring in $AdS_3 \times S^3$
 with a R-R 3-form  background. 
In Section 6 we obtain a realization of the generators  of the 
basic symmetry 
 superalgebra 
$psu(2,2|4)$ as Noether charges expressed 
 in terms of  2-d fields which are 
the coordinates of the  $AdS_5 \times S^5$ superstring 
 in the light-cone gauge.

We collect various technical details in four appendices.
In Appendix A we summarize the notation used in this paper.
In Appendix B we describe some  relations between
two  different ``5+5"
parametrizations of $S^5$ (in terms of 5 Cartesian  coordinates
and in terms of  unit 6-d vector) which  are useful
for translation between the corresponding forms of the
\adss  superstring action.
In Appendix C  we  review  the  two forms of 
the $\k$-symmetry \lc
gauge fixed  action  which 
use the ``4+6" parametrizations of \adss \ci{MT3} and 
write down  the corresponding  phase space  Lagrangians.
 In Appendix D  we present some details of the construction
of conformal supercharges in Section 6.

%%%%%%%%%%%%%%%%%%%%%%%%%%%%%%%%%%%%%%%
\newsection{Bosonic string in curved space:
light-cone gauge approach}
%%%%%%%%%%%%%%%%%%%%%%%%%%%%%%%%%%%%%%%%%%%%%%%

 Let us start with a review of some 
 previous discussions 
 of   \lc gauge fixing   for bosonic strings 
 in curved space. 
   In flat space in  BDHP formulation \ci{bdh,poly}
 one starts by  fixing  the
 conformal gauge
\be
 \g^{\vm\vn}   = \eta^{\vm\vn}\ , \qquad \ \ \ \
  \g^{\vm\vn}\equiv \sqrt{g}  g^{\vm\vn} \ ,  \ \ \ \  
  g\equiv -\det g^{\vm\vn}\ , 
  \ \ \ \  
  \det \g^{\vm\vn}  =-1
   \ , \la{coonf}\ee
     and then  fixes the residual conformal diffeomorphism
   symmetry
   on the plane by choosing $x^+(\tau,\sigma)  = \tau$.\foot{For a
   discussion of various ways
  of fixing the \lc gauge   in the case of flat target
  space  see, e.g., \ci{smitze}.}
 An  alternative (equivalent) approach
  is   to use  the original  GGRT \ci{ggrt}
 formulation  based on writing the Nambu action in the
   first order form   and fixing the
  diffeomorphisms  by the two   conditions -- on  one coordinate
  and on one canonical
  momentum:
\be 
x^+ = \tau\ , \ \ \ \ \ \ \ \ {\cal P}^+  =\const \ . \la{ggrt}
\ee
The obvious  requirement for being  able to choose 
the \lc gauge in a  curved space is 
the existence of a null Killing vector.
The  first  approach  based on  fixing  the conformal gauge
for the 2-d metric 
does not in general  apply in curved spaces
with null Killing vectors which are not of the direct
product
form $R^{1,1} \times M^{d-2}$. The exception is the case when 
 the null Killing vector is covariantly constant
\ci{hors}.  One  is thus forced to   give up 
 the standard conformal gauge   \rf{coonf} and 
  fix the diffeomorphisms
  by  imposing instead, e.g., 
 $\g^{00}=-1, \   x^+ =  \tau$. This  gauge choice   is
 consistent provided the background metric 
 satisfies
$G_{+-}=1, G_{--}=G_{-\al }=0,\ \del_- G_{mn} =0$
\ci{rudd}. 

 The above  conditions do not apply in the case of 
the $AdS$-type metric ($x_\al$ stands for all
$d-3$ transverse coordinates):

\be
ds^2 =  G(\p) dx^a dx^a  + d\p^2=  
 G(\p) (-dt^2 +  dx^2_\al + dx^2_3 ) +  d\p^2 \ . \la{mettr}
\ee
Indeed, the 
two   null Killing vectors  here are  not covariantly constant
 and  also $G_{+-} = G(\p)  \not=1$.
However, 
  a slight modification    of the above 
  gauge conditions of \ci{rudd}
on   $\g^{00}, \ x^+$
 represents  a consistent  gauge choice 
 \ci{MT3}\foot{A closely related 
\lc gauge choice $\g^{\mu\nu}=\diag( - G^{-1}, G), \
x^+=\tau$ 
 was originally suggested by A. Polyakov \ci{pol}.}

  \be
G(\p) \g^{00}=-1\ , \   \ \ \ \ \ \ \ x^+ =  \tau  \ .
 \la{our}
 \ee 
  There is  a potential
  complication 
that the norm of the Killing vector  may  vanish  if there is
a  horizon  where $G=0$. We shall adopt a formal approach,  
assuming that the  degeneracy of the 
\lc  description  reflected in the
$G\to 0$ singularity  of the resulting \lc gauge fixed action 
should have some    physical resolution, e.g.,  the form of the 
\lc  Hamiltonian  may  suggest how  the wave functions 
 should be defined in this region.

The case of \rf{mettr}  with $G= e^{2\p}$ corresponds to 
the $AdS$ space  in a Poincar\'e  coordinate patch.
One needs to use the Poincar\'e coordinates to have a null
 isometry in the bulk  and  at the boundary (the boundary
 should have  $R^{1,3}$ topology).\footnote{The  Poincar\'e parametrization
 was  used  in the light-cone formulation of
field dynamics in $AdS$ spacetime \ci{met1,met2}.}
 %Note that light-cone formalism in AdS
%spacetime (see \ci{met1,met2}) as compared to the one in flat spacetime
%keeps manifest invariance with respect of only $so(d-3)$ algebra}):
The      AdS/CFT duality suggests that 
since  the boundary SYM theory 
in $R^{1,3}$  has a well-defined light-cone gauge 
description \ci{brink}, it should be possible to fix 
 some analog of a \lc gauge  for the dual string theory as well. 
This is the motivation  behind our \lc gauge 
fixing approach, in spite of
the fact that  there is no
 globally well-defined null Killing  vector  in $AdS$ space
 (its  norm proportional to $e^{2\phi}$
 vanishes at the horizon $\phi=-\infty$ of the  Poincar\'e 
 patch).\foot{This subtlety  and the issue of 
  fixing  a global diffeomorphism gauge 
for  $AdS$ string  was discussed in \ci{roz}.}

Some  further comments on the 
  coordinate space BDHP  approach using  \rf{our} 
  may be found in \ci{MT3}.
  Here we shall follow the  
 phase space  GGRT  approach \ci{ggrt}  based on  fixing the
 diffeomorphisms in curved space  by
   the {\it same}  gauge
 \rf{ggrt} as in flat space.
 Starting with the string  Lagrangian corresponding to 
 \rf{mettr} (we set $2\pi \a'=1$) 

 \be
{\cal L} = -   \ha h^{\vm\vn} \left[  \del_\vm x^a \del_\vn x^a 
 + G^{-1}(\p)\del_\vm \p\del_\vn \p\right] \ , 
 \qquad
 h^{\vm\vn} \equiv \g^{\vm\vn}G(\p) \ , 
 \la{lll}
 \ee
we get the canonical momenta  for $x^a=(x^+,x^-,x^\al)$ and $\p$
(dot and prime are derivatives over $\tau$ and $\sigma$,
 see Appendix A for  definitions)
\be
\PP_a = -  h^{00}\dot{x}_a - h^{01} \x'_a \ , 
\ \ \ \  
\Pi = -  h^{00}G^{-1}\dot \p - h^{01}G^{-1} \ph' \ . 
\la{ppp}
\ee
The phase space Lagrangian is then  

\begin{eqnarray} 
{\cal L} &=& \dot x_\al  \PP_\al   + \dot \p \Pi + \dot x^+ \PP^-  +
\dot x^- \PP^+   
\nonumber\\ 
&+& \  { 1 \ov 2 h^{00}} 
\bigg[(  \PP^2_\al + 
2\PP^+\PP^-) + G^2(\p)  (  \x'^2_\al  +  2\x'^+ \x'^-) + G(\p)(\Pi^2 + \ph'^2 ) 
\bigg]
\la{logj} \nonumber \\ &+& \  { h^{01} \ov  h^{00} }( \x'_\al \PP_\al    
+ \ph'\Pi  + \x'^+ \PP^-  + \x'^- \PP^+) \ , 
\end{eqnarray} 
where $ 1/h^{00}$ and  ${ h^{01}/h^{00} }$ play the role of the Lagrange
multipliers. 
As in flat space, $x^+$  is a natural choice for the 
evolution parameter.\foot{In \ci{met1,met2} it was demonstrated that 
in field theory  in $AdS$ space in  Poincar\'e coordinates 
 $x^+$ can be treated 
as a  light-cone evolution parameter. Therefore,  it is 
reasonable to consider $x^+$  as a light-cone 
evolution parameter in  $AdS$  string theory too.}  
This suggests choosing the \lc gauge  as in \rf{ggrt} 

\be 
  x^+ = \tau\ , \ \ \ \ \ \ \ \ \PP^+  = p^+ l\inv = \const 
   \ .  \la{grt}
\ee  
For flexibility,  
we introduced the length $l$ of the string 
parameter $\s$.  $l$ may be chosen to be equal to $p^+$ 
(which is useful in the study of interactions) or to $
\sqrt{ 2 \pi \a'}=1$  (which is useful in the study of
 the spectrum).
 In Section 3 we shall set   $l=p^+$, while in 
  Sections 4-6 we shall use $l=1$.
 To transform the expressions in  Sections 4-6
 into the ``$l=p^+$'' scheme ($\PP^+=1$) 
 one is simply to put $p^+=1$  in the Lagrangian 
 and Hamiltonian densities.

Integrating over $\PP^-$ we get the relation 

\be
h^{00}    = - p^+ l^{-1} \ . \la{guu}
\ee
%%NEW%%
Note that this is equivalent (for $l=p^+$)
to the gauge condition on the 2-d metric in \rf{our}.
The expression for $\PP^-$ follows from the $1/h^{00}$
constraint.
Integration  over the non-constant part of $  { h^{01}/h^{00} }$ 
leads to 

\be
p^+ l\inv  \x'^-   +  \x'_\al  \PP_\al    + \ph' \Pi =0  \ , \la{conn}
\ee
and   integration over its constant  part  gives
\be
\int^{l}_0 d\s ( \x'_\al  \PP_\al   + \ph' \Pi) =0   \ . \la{intconn}
\ee
The resulting \lc Hamiltonian  is given (up to the sign, 
$H=-P^- = - \int d \sigma \PP^-$)  by  
\be
P^- = -{ l \ov 2 p^+ } 
 \int^{l}_0 d\s \bigg[  \PP^2_\al  + G^2(\p)  \x'^2_\al 
 + G(\p) (\Pi^2 +  \ph'^2)  \bigg] \ . 
\la{hamm}
\ee
Note that the equation for $x^-$ implies  that 
$ ({ h^{01}/h^{00} })'=0$. 
%so that ${ h^{01}} =0$ 
Since it also implies ${ h^{01}} =0$ at open string ends, it
follows that 
${ h^{01}} =0$ in the case of open string. 
%and $ {h^{01} } =$const in the case of the 
But we can only conclude that $ {h^{01} } =$const in the case of
closed string. This constant Lagrange multiplier imposes
the integrated diffeomorphism constraint \rf{intconn}, necessary
for consistency in the closed string case.

Let us note that fixing the light-cone gauge in the action, before
obtaining the equations of motion, results in lost equations
of motion which would be obtained by varying $x^+$ and ${\cal P}^+$.
It is easy to check that these equations (obtained before
gauge fixing) determine ${\dot x}^-$ and
$\dot{\cal P}^-$, the time derivatives of variables that
have been eliminated by the diffeomorphism constraints.
Indeed, the expressions they give for  $\dot{\cal P}^-$
and $\partial_\sigma{\dot x}^-$ are equivalent to
ones obtained by differentiating the
constraints with respect  to time. Thus the only new information
contained in the lost equations is information about the zero mode of
$x^-$, namely,  
$\int_0^l d\sigma{\dot x}^-={l\over p^+}P^- .$
Since $x^-$ does not enter the gauge-fixed
Hamiltonian this new information is unnecessary for subsequent
analysis of the dynamics of the system.

%%%NEW%%%
The bosonic string in $AdS$ space is not consistent (conformally
 invariant) at the quantum level.  To make it 
 conformally invariant one may  add extra couplings to an NS-NS 
 3-form  background or add fermions and 
  consider superstring coupled to an R-R 
 background. The simplest example is  bosonic string 
 propagating on the $SL(2,R)$ group  manifold described 
 by the  WZW model. It is instructive to
 demonstrate  how the phase space \lc gauge approach 
 described above applies in this case.
 In the standard Gauss
 parametrization,  the Lagrangian of the $SL(2,R)$
 WZW  model is the same as \rf{lll} (with $G= e^{2 \p}$)
 plus a WZ term

\be
{\cal L}
=-\sqrt g g^{\vm\vn}(e^{2\phi}\partial_\vm x^+\partial_\vn  x^-
 +\frac{1}{2}\partial_\vm \phi \partial_\vn \phi)
+\epsilon^{\vm\vn}e^{2\phi}\partial_\vm x^+\partial_\vn x^-
\ . \ee
The corresponding phase space
Lagrangian is  found to be    ($ h^{\vm\vn}
\equiv  \sqrt g g^{\vm\vn}e^{2\phi}$) 
\begin{eqnarray}
{\cal L}
&=&
 \PP^+\dot{x}^- + \PP^-\dot{x}^+ +  \Pi\dot{\phi}
+\frac{1}{2h^{00}}\Bigl[2\PP^+\PP^- 
+e^{2\phi}(2\PP^- \x'^+  -  2\PP^+ \x'^-
+\Pi^2+\ph'^2)\Bigl]
\nonumber\\
&+&
\frac{h^{01}}{h^{00}}(\PP^+\x'^- + \PP^-\x'^+ +  \Pi\ph')\ . 
\la{wzwe}
\end{eqnarray}
Compared to the $AdS_d$ model \rf{logj}
in the present $d=3$ case 
there is no  transverse degrees of freedom,
 and  because of the  WZ coupling  
 the Lagrangian \rf{wzwe}
does not contain  $\x'^+\x'^-$ term.
 Fixing the \lc gauge as in \rf{grt}  and integrating over
  ${\cal P}^-$  (getting again \rf{guu})
  we  find  ($l=1$) 
\be
{\cal L}
=\Pi\dot{\phi}
-\frac{e^{2\phi}}{2p^+}(\Pi^2 + \ph'^2)+ e^{2\phi}\x'^-
-\frac{h^{01}}{p^+}(p^+\x'^- +  \Pi\ph')\ . 
\ee
Using the expression for  $\x'^-$ which is implied by 
the constraint  we  can rewrite 
this   Lagrangian as  

\be
{\cal L}
=\Pi\dot{\phi}+\PP^- -\frac{h^{01}}{p^+}(p^+\x'^- +  \Pi\ph')
\ , \ee
where the (minus)  Hamiltonian density   
is

\be
\PP^- = 
-\frac{e^{2\phi}}{2p^+}(\Pi + \ph')^2 \ . 
\ee
This is to be compared with 
the pure metric  case \rf{hamm} which for 
 the $AdS_3$  gives
 \be
 \PP^- = 
-\frac{e^{2\phi}}{2p^+}(\Pi^2 + \ph'^2) \ . \la{addss}
\ee

%%%%%%%%%%%%%%%%%%%%%%%%%%%%%%%%%%%%%%%%%%%
\newsection{Some classical string solutions}
%applications}
%%%%%%%%%%%%%%%%%%%%%%%%%%%%%%

To illustrate the utility of the bosonic \lc Hamiltonian 
derived in Section 2, we shall  demonstrate 
 how the (classical)   discussion  of the  simplest 
Wilson loops  in \adss 
(straight line and parallel lines) given in \ci{malda}
can be phrased in the present \lc gauge setting. 
Since we shall  consider  only the classical string approximation 
  it is sufficient to  ignore the fermions, i.e. to start 
 with the bosonic \lc Hamiltonian \rf{hamm}
 (here we set  $l=p^+$)
% rescale the variables to  restore
% the dependence on $R$ and $\a'$) 
\begin{eqnarray}
H={1\over2}\int_{0}^{p^+} d\sigma\left[{\PP}^2_\al 
+ e^{2\phi/\gamma }\x'^2_\al 
+e^{\phi/\gamma}(\Pi^2+\ph'^2)\right] \ , 
\end{eqnarray}
where $4\gamma^2=R^2T_0={R^2/ 2\pi\a^\prime}$.

%%%%%%%%%%%%%%%%%%%%%%%%%%%%%%%%
\subsection{ Straight string: isolated ``quark" source}
%%%%%%%%%%%%%%%%%%%%%%

A quark source is represented as   a static open string stretched from the
horizon $\phi=-\infty$ to the boundary $\phi=+\infty$
of $AdS$ space.
For static solutions, ${\PP}_\al=\Pi=0$ and the classical
equations reduce to those extremizing the Hamiltonian:
\begin{eqnarray}
(e^{2\phi/\gamma}\x'_\al)^\prime&=&0 \ , \\
{1\over\gamma}e^{2\phi/\gamma}\x'^2_\al
+{1\over2\gamma}e^{\phi/\gamma}\ph'^2
-(e^{\phi/\gamma}\ph')^\prime&=&0 \ . \la{sec}
\end{eqnarray}
The first equation implies
\begin{eqnarray}
{\x'_\al}={T_\al }e^{-2\phi/\gamma}\ ,
\end{eqnarray}
where ${T_\al }$ is an integration constant. Note that for 
static solutions, the constraint \rf{conn}, i.e. 
$-\x'^-={\x'_\al} {\PP}_\al
+\ph' \Pi$,  implies that there is no extension in $x^-$.
For an isolated quark, we also want no extension in ${x_\al}$,
so, in this case, we have ${T_\al }=0$.

If we define $\rho=e^{\phi/\gamma}$, the second equation \rf{sec} 
implies
\begin{eqnarray}
{\rh'^2\over2\rho}+{T^2_\al \over2\gamma^2\rho^2}=C\ ,
\end{eqnarray}
where  $C$ is  a  constant of integration. The left hand side
is $1/\gamma^2$ times the density of $P^-$, so we may
identify $C=p^-/\gamma^2p^+$. For $T_\al=0$, one can trivially
integrate the equation to determine
\begin{eqnarray}
\sqrt{\rho}={\sigma\over2\gamma}\sqrt{|{2p^-\over
p^+}|}\ ,
\end{eqnarray}
where we have fixed the integration constant so that the
end at $\sigma=0$ is on the horizon ($\rho=0$). The other end
is at $\sigma=p^+$, so we find
\begin{eqnarray}
m_q\equiv\sqrt{2|p^+p^-|}=2\gamma\sqrt{\rho_{\rm max}}\ .
\end{eqnarray}
If that end reaches the boundary, $\rho_{\rm max}\to\infty$, 
implying an infinite mass for the quark source, to be expected
for a point charge.

%%%%%%%%%%%%%%%%%%%%%%%%%%%%%
\subsection{String ending on the boundary:
``quark-antiquark" source at separation $L$}
%%%%%%%%%%%%%%%%%%%%%%

A quark-antiquark source 
corresponds to an open string with both ends on the boundary
separated by the distance $L$. In this case $T_\al \neq0$ and there
is a minimum value $\rho_{\rm min}>0$ of $\rho$ for an
interior point of the string. For reasons of symmetry we shift
the $\sigma$ range to mark this point by $\sigma=0$:
$-{p^+/ 2} <\sigma<{p^+/2} $. $\rho_{\rm min}$ occurs at the
point where $\rh'=0$, so $C={
T^2_\al/ 2\gamma^2\rho_{\rm min}^2} $. 
Thus we have
\begin{eqnarray}
m_{q{\bar q}}\equiv\sqrt{2|p^+ p^-|}={T_\al p^+\over\rho_{\rm min}}\ .
\end{eqnarray}
The separation between the ends of string is obtained by
integrating ${\x'_\al}$:
\begin{eqnarray}
L\equiv\left|\int d\sigma\ \x'_\al\right|
&=&T_\al\int_{-{p^+\ov 2} }^{{p^+\ov 2} }{d\sigma\over\rho^2}
=2\gamma\int_{\rho_{\rm min}}^{\rho_{\rm max}}{d\chi\over\chi^2
\sqrt{\chi/\rho_{\rm min}^2-1/\chi}}\nonumber
\\
&=&{2\gamma\over\sqrt{\rho_{\rm min}}}\int_{1}^{\rho_{\rm
max}/\rho_{\rm min}}
{d\chi\over\chi^2
\sqrt{\chi-1/\chi}}\to{4\gamma\over\sqrt{\rho_{\rm min}}}
{\Gamma(3/4)\Gamma(1/2)\over\Gamma(1/4)},
\end{eqnarray}
where the last form is the limit as $\rho_{\rm max}\to\infty$.
Similarly, we obtain $\sigma$ as a function of $\rho$ by direct
integration
leading to
\begin{eqnarray}
{p^+\over2}&=&{\gamma(\rho_{\rm min})^{3/2}\over T_\al}
\int_{1}^{\rho_{\rm max}/\rho_{\rm min}}
{d\chi\over\sqrt{\chi-1/\chi}}\ , \\
\sqrt{2|p^+p^-|}&=&2\gamma\sqrt{\rho_{\rm min}}
\int_{1}^{\rho_{\rm max}/\rho_{\rm min}}
{d\chi\over\sqrt{\chi-1/\chi}}\nonumber \\
&\to&4\gamma\sqrt{\rho_{\rm max}}-4\gamma\sqrt{\rho_{\rm min}}
{\Gamma(3/4)\Gamma(1/2)\over\Gamma(1/4)}\ ,
\end{eqnarray}
where the last form gives the non-vanishing terms of the
behavior as $\rho_{\rm max}\to\infty$. We finally eliminate
$\rho_{\rm min}$ in favor of $L$ to reach the final result:
\begin{eqnarray}
\sqrt{2|p^+p^-|}
&\to&4\gamma\sqrt{\rho_{\rm max}}-{16\gamma^2\over L}
{\Gamma(3/4)^2\Gamma(1/2)^2\over\Gamma(1/4)^2}\nonumber\\
&=&4\gamma\sqrt{\rho_{\rm max}}-{4\gamma^2\over L}
{(2\pi)^3\over\Gamma(1/4)^4}.
\end{eqnarray}
To compare with the known result \ci{malda}, recall that 
$4\gamma^2=R^2T_0=R^2/2\pi\a^\prime=\sqrt{\lambda}/2\pi$,
where $\l$ is the  `t Hooft coupling.
The first divergent term is just twice the isolated quark source
mass,
so the second finite term is the predicted interaction energy 
between quark and antiquark.

%%%%%%%%%%%%%%%%%%%%%%%%%%%%%%%%%%%%%%%%%%
\newsection{  Superparticle in $AdS_5\times S^5$: 
light-cone  Hamiltonian
}
%%%%%%%%%%%%%%%%%%%%%%%%%%%%%%

Before discussing superstring
 it is instructive   to consider  first a 
superparticle in $AdS_5 \times S^5$ space.
The covariant  $\kappa$-symmetric  action for a superparticle 
in $AdS_5\times S^5$ can be obtained from the superstring action of 
\ci{MT1} 
by taking the zero slope  limit $\alpha^\prime \rightarrow 0$.
By applying  the \lc gauge fixing 
procedure described  here 
 one could then  obtain the 
superparticle light-cone gauge fixed action. 
One the other hand,  there is  a method \ci{dir}
which reduces the problem of
finding a new  (light-cone gauge) 
 dynamical system  to the problem of finding a new
solution of the commutation relations of the  defining symmetry algebra 
(in our case  $psu(2,2|4)$ superalgebra).
This method was applied to the  $AdS$ superparticle case 
in \ci{met3}. In the  notation
 of the present  paper the 
quantum (operator-ordered)    light-cone  Hamiltonian
($H= -\PP^-$)  for  the superparticle
 found there  has  the following form
\be\label{parham}
\PP^-
= -\frac{1}{2p^+}\Bigl(\PP_\al^2
+e^{\phi}\Pi e^\phi \Pi +e^{2\phi} A\Bigr) , 
\ee
\be\la{ax}
A \equiv X - \frac{1}{4}\,,
\qquad
X\equiv 
l^i{}_j^{\,\,2} +(p^+\eta^2-2)^2 + 4p^+\eta_i l^i{}_j \eta^j \ , 
\ee
where
$
\eta^2\equiv \eta^i\eta_i
$
and $l^i{}_j$ is  the angular momentum operator of the $su(4)$ algebra 
(for details see  \ci{met3}).
The $\PP_\al= (\PP,\bar{\PP})$ and $\Pi$ are the bosonic conjugate momentum 
 operators as in \rf{logj}.
The odd part of the  phase space is represented by 
 $\theta^i$, $\eta^i$ considered as fermionic 
coordinates and $\theta_i,$ $\eta_i$ considered as fermionic momenta. 
Note that the  Hamiltonian does not depend on the fermionic
variables $\theta^i$ and
$\theta_i$ 
(present  in the  light-cone gauge 
formulation of 4-dimensional  $\N=4$  SYM theory  \ci{brink})
 but 
they  will  appear in
the phase space 
Lagrangian as   $\theta^i \dot{\theta}_i + \theta_i \dot{\theta}^i $.

 The canonical operator commutation relations 
 are\foot{Note that in  the rest
 of the paper the brackets will stand for the classical Poisson
 brackets, and thus there will be no i's in similar relations.}

\be
[\PP,\bar{x}]=-{\rm i}\,,
\qquad
[\bar{\PP},x]=-{\rm i}\,,
\qquad
[\Pi,\phi]=-{\rm i}\,,
\ee
\be
\{\theta^i,\theta_j\}=\frac{1}{p^+}\delta_j^i\ , \ \ 
\qquad
\{\eta^i,\eta_j\}=\frac{1}{p^+}\delta_j^i\,.
\ee
The operator $A$  is equal to zero only for massless 
representations 
which are  irreducible representations of the conformal algebra
\cite{met5,met2} ($so(5,2)$ in  the case of $AdS_5$ space). 
The important property  of the operator $X$ (\ref{ax}) is that 
its eigenvalues  are equal to squares of  integers 
for all states of type IIB supergravity 
compactified on $S^5$. 
This fact plays an important role in formulating the  AdS/CFT
correspondence  for chiral primary states \ci{met3}. The relation
(\ref{ax}) then implies that the  operator $A$ is
never equal to zero and thus  
the  scalar fields \cite{krn} as well as all other 
modes  \ci{met3} of $S^5$  compactified  type IIB  supergravity 
have equations of motion which are not
 conformally invariant.
We expect that this  property of the operator $X$  should have a string-theory 
analog.

The \lc gauge phase space 
Lagrangian  for the superparticle  in \adss    is obtained  from the 
Hamiltonian \rf{parham} in the usual way

\be
{\cal L} = 
\PP_\al\dot{x}_\al +  \Pi\dot{\phi} 
+ \PP_M \dot{u}^M
+\frac{{\rm i}}{2}p^+(\theta^i \dot{\theta}_i
+\eta^i\dot{\eta}_i+\theta_i
\dot{\theta}^i+\eta_i\dot{\eta}^i)
+\PP^- \ . 
\ee
$u^M$ is a unit  6-d  vector  used to parametrize 
$S^5$. Note that here (and  in the  string case) we treat 
 $x^-$, $\PP^+=p^+$ separately from the rest
of the phase space variables; 
 $p^+$ is conserved,  while $x^-$ satisfies the equation 
\be
\dot{x}^- = \frac{1}{p^+}\PP^-\,.
\la{marrr}
\ee

%%%%%%%%%%%%%%%%%%%%%%%%%%%%%%%%%%%%%%%%%%
\newsection{Light-cone  Hamiltonian approach to 
 superstring in \adss}
%%%%%%%%%%%%%%%%%%%%%%%%%%%%%%

%%%%%%%%%%%%%%%%%%%%%%%%%%%%%%%%%%%%%
\subsection{Review of  $\k$-symmetry  gauge fixed  action}
%%%%%%%%%%%%%%%%%%%%%%%%

Let us first recall 
 the  form of the  superstring action in the 
 $\k$-symmetry light-cone gauge
  \ci{MT3}. It  is formulated in
terms of 10 bosonic coordinates $(x^\pm,x,\bar{x};\phi,y^\sca)$
($y^\sca$  are 5 independent coordinates of $S^5$)  
and 16 
fermionic
coordinates $(\theta^i,\theta_i,\eta^i,\eta_i)$ which transform 
in the fundamental
representations of $SU(4)$.
 The Lagrangian (equivalent to the one in \rf{actki})
 is  given by the sum of the ``kinetic" and ``Wess-Zumino" terms 
(see Appendices  A and B  for notation) 

$$
{\cal L}= {\cal L}_{kin}+{\cal L}_{WZ}\ , 
$$
$$
{\cal L}_{kin}
=
-\sqrt{g}g^{\vm\vn}\Bigl[
e^{2\phi}(\partial_\vm x^+ \partial_\vn x ^-
+ \partial_\vm x\partial_\vn\bar{x})
+\frac{1}{2}\partial_\vm \phi\partial_\vn \phi
+\frac{1}{2}G_{\sca\scb}(y) D_\vm y^\sca D_\vn y^\scb\Bigr]
$$
\begin{equation}
- \ \frac{{\rm i}}{2} \sqrt{g}g^{\vm\vn}
e^{2\phi}\partial_\vm x^+
\Bigl[\theta^i\partial_\vn \theta_i
+\theta_i\partial_\vn \theta^i
+\eta^i\partial_\vn \eta_i
+\eta_i\partial_\vn \eta^i 
+{\rm i}  e^{2\phi}\partial_\vn x^+(\eta^2)^2\Bigr]\ ,
\label{actkin3}
\end{equation}

\begin{equation}\label{actwz3}
{\cal L}_{WZ}
=\epsilon^{\vm\vn}
e^{2\phi}\partial_\vm x^+ \eta^i C_{ij}^U
(\partial_\vn\theta^j-{\rm i}\sqrt{2}e^\phi \eta^j
\partial_\vn x)+h.c. \ , 
\end{equation}
where 
\begin{equation}\label{dyan}
D_\mu  y^\sca = \partial_\mu  y^\sca - 2{\rm i}\eta_i
(V^\sca)^i{}_j\eta^j e^{2\phi} \partial_\mu  x^+  \ .
\end{equation}
Here  $G_{\sca\scb}$ 
and  $(V^\sca)^i{}_j$ are the metric tensor and the 
Killing vectors of $S^5$ respectively, i.e.
this  Lagrangian  corresponds to the following 
parametrization  of  the  metric of
$AdS_5\times S^5$

\be
ds^2 =e^{2\phi}dx^adx^a +d\phi^2+ G_{\sca\scb}dy^\sca dy^\scb \ . 
\la{ded}
\ee
This  form of the superstring action  (which we shall call ``intermediate")
is most convenient for  deriving 
 other  forms which differ by the way one chooses the bosonic 
 coordinates that parametrize  \adss (see \ci{MT3}
and Appendices B and C).
 
 Another useful form is found by using a  6-d unit vector  $u^M$ 
 to parametrize $S^5$, i.e. by 
replacing \rf{ded} by 
\be
ds^2 =e^{2\phi}dx^adx^a +d\phi^2+ du^M du^M\ , \ \ \  \ \ \ \ \ \   u^M u^M=1 \ .
\la{uuu}
\ee
 Then \rf{actkin3},\rf{actwz3} are replaced by 
$$
{\cal L}_{kin}
=
-\sqrt{g}g^{\vm\vn}\Bigl[
e^{2\phi}(\partial_\vm x^+ \partial_\vn x ^-
+ \partial_\vm x\partial_\vn\bar{x})
+\frac{1}{2}\partial_\vm \phi\partial_\vn \phi
+\frac{1}{2} D_\vm u^M D_\vn u^M\Bigr]
$$
\begin{equation}
- \ \frac{{\rm i}}{2} \sqrt{g}g^{\vm\vn}
e^{2\phi}\partial_\vm x^+
\Bigl[\theta^i\partial_\vn \theta_i
+\theta_i\partial_\vn \theta^i
+\eta^i\partial_\vn \eta_i
+\eta_i\partial_\vn \eta^i 
+{\rm i}  e^{2\phi}\partial_\vn x^+(\eta^2)^2\Bigr]\ ,
\label{actkin6}
\end{equation}

\begin{equation}\label{actwz6}
{\cal L}_{WZ}
=\epsilon^{\vm\vn}
e^{2\phi}\partial_\vm x^+ \eta^i \rho_{ij}^M u^M
(\partial_\vn\theta^j-{\rm i}\sqrt{2}e^\phi \eta^j
\partial_\vn x)+h.c.\ , 
\end{equation}
where 

\be
D_\vm u^M = \partial_\vm u^M 
-2{\rm i}\eta_i (R^M)^i{}_j\eta^j e^{2\phi}\partial_\vm x^+ \ , \ \ \ \  \ \ \ \  
R^M = - { 1 \ov 2} \rho^{MN} u^N  \ . 
\ee
The parametrization  using $u^M$  is the most convenient one 
 for the  discussion
of the superparticle in $AdS_5\times S^5$ \ci{met3,met4}, 
and is  well-suited  for the harmonic  decomposition of  the 
light-cone superfield of type IIB supergravity into the  Kaluza-Klein modes
\ci{met3}.
%It should be useful also  in the superstring case.
Other  forms  of the superstring action 
 using 4+6 Cartesian  coordinates  for \adss 
 (see \rf{actki} and Appendix C) are directly 
 related to \rf{actkin6},\rf{actwz6}  by a coordinate 
 transformation ($Y^M= e^{\p} u^M$, $|Y|= e^{\p}$).

The  superstring Lagrangian (\ref{actkin3}),(\ref{actwz3}) 
and all of its  other  forms  mentioned above 
can be represented   in the following  way

\be\la{lagdec}
{\cal L}= {\cal L}_1+{\cal L}_2 +{\cal L}_3 \    ,   
\ee
where the three parts are 

\be
{\cal L}_1\label{l1}
= -h^{\vm\vn}\partial_\vm x^+\partial_\vn  x^-
+\partial_\vm x^+ A^\vm 
+\frac{1}{2}h^{\vm\vn}\partial_\vm x^+\partial_\vn  x^+ B
-\frac{1}{2}h^{\vm\vn} g_{\sca\scb}D_\vm y^\sca D_\vn y^\scb \ , 
\ee

\be\label{l2}
{\cal L}_2= 
-\frac{1}{2}h^{\vm\vn}\partial_\vm x^\al \partial_\vn  x^\al
+\partial_\vm x^\al C^{\vm \al}\ , 
\ee

\be\label{l3}
{\cal L}_3 = -\frac{1}{2}h^{\vm\vn}e^{-2\phi}
\partial_\vm \phi \partial_\vn \phi + T\   . 
\ee
Here  $x^\al = (x, \bar x)$ , 
\be 
g_{\sca\scb}\equiv  e^{-2\phi}G_{\sca\scb} \ ,  \ \ \ \ \ \ \ \ \ 
D_\vm  y^\sca \equiv \partial_\vm y^{\sca} + F^\sca \partial_\vm x^+ \  ,  
\ee
and  $h^{\vm\vn}$ is defined as in (\ref{lll}), i.e. 
\be \la{hhh}
h^{\vm\vn}\equiv \sqrt{g}g^{\vm\vn}e^{2\phi}\,,
\qquad
h^{00}h^{11}-(h^{01})^2=-e^{4\phi} \ . 
\ee
The decomposition (\ref{lagdec}) is made so that  
the functions $A^\vm$, $B$, $C^{\vm\al}$, $F^\sca$ depend  on 
(i)  the anticommuting  coordinates and their derivatives with respect to
 both  worldsheet coordinates  $\tau$ and $\sigma$, and  
(ii)  the  bosonic coordinates 
and their derivatives with respect to the world sheet spatial 
coordinate $\sigma$ only. 
The reason for  this   decomposition is that  we shall  use  the 
 phase space description    with respect to the 
bosonic coordinates only,  i.e. we  shall not make the Legendre transformation
with respect to the fermionic coordinates.

In  the case of  the ``intermediate"  form  of the action 
(\ref{actkin3}),(\ref{actwz3}) these  functions take the following form
\begin{equation}\label{funA}
A^\vm
=-\frac{{\rm i}}{2}h^{\vm\vn}(\theta^i\partial_\vn \theta_i
+\eta^i\partial_\vn \eta_i)
+\epsilon^{\vm 1}
e^{2\phi}\eta^i C_{ij}^U
(\th'^j - {\rm i}\sqrt{2}e^\phi\eta^j\x')+h.c. \ , 
\end{equation}
\begin{eqnarray}
\label{funB}&&
B
= e^{2\phi}(\eta^2)^2\ ,
\\
\label{funC}
&&
C^{\vm \bar{x}}
=
{\rm i}\sqrt{2}\epsilon^{\vm 1}
e^{3\phi}\x'^+ \eta^i C_{ij}^U\eta^j\,,
\qquad
C^{\vm x} = (C^{\vm \bar{x}})^\dagger\ , 
\\
\label{funF}
&&
F^\sca 
= -2{\rm i}e^{2\phi}\eta_i (V^\sca)^i{}_j\eta^j\,,
\\
\label{funT}
&&
T
=-e^{2\phi}\x'^+ \eta^i C_{ij}^U\dot{\theta}^j +h.c.\,.
\end{eqnarray}

%%%%%%%%%%%%%%%%%%%%%%%%%%%%%%%%%%%%%
\subsection{Phase space Lagrangian}
%%%%%%%%%%%%%%%%%%%%%%%%

Computing  the canonical momenta for the bosonic coordinates 

\be
\PP_a 
= \frac{\partial {\cal L}}{\partial \dot{x}^a}\,,
\qquad
\Pi
= \frac{\partial {\cal L}}{\partial \dot{\phi}}\,,
\qquad
\PP_\sca 
= \frac{\partial {\cal L}}{\partial \dot{y}^\sca}\,,
\ee
we get 
\begin{eqnarray}
\label{can1}
\Pi 
&=& -h^{00}e^{-2\phi}\dot{\phi}^+ - h^{01}e^{-2\phi}\ph'^+\ , 
\\
\PP^+ 
&=& -h^{00}\dot{x}^+ -h^{01}\x'^+\ , 
\\
\PP^\al
&=&
-h^{00}\dot{x}^\al -h^{01}\x'^\al +C^{0\al}\ ,
\\
\PP^\sca 
&=&
-h^{00}\dot{y}^\sca
-h^{01}\y'^\sca +F^\sca \PP^+\ , 
\\
\label{can6}
\PP^- 
&=&
-h^{00}\dot{x}^- - h^{01}\x'^- +A^0 - B\PP^+ +\PP_\sca F^\sca \ . 
\end{eqnarray}
where $\PP^\pm \equiv \PP_\mp$, $\PP^\sca\equiv g^{\sca\scb}
\PP_\scb$. 
By applying the same  procedure as in the bosonic case 
we find then  the following 
phase space Lagrangian ${\cal L}= {\cal L}_1 + {\cal L}_2 + {\cal L}_3$
(cf. \rf{logj}) 

\begin{eqnarray}
{\cal L}_1
&=&
 \PP^+\dot{x}^- + \PP^-\dot{x}^+ + \PP_\sca \dot{y}^\sca
+\frac{1}{2h^{00}}\Bigl[2\PP^+\PP^- + 2e^{4\phi}\x'^+\x'^-
\nonumber
\\
&+&
g^{\sca\scb}\PP_\sca \PP_\scb
+e^{4\phi}g_{\sca\scb}D_1y^\sca D_1 y^\scb
+(\PP^{+2}-e^{4\phi}\x'^{+2})B - 2F^\sca \PP_\sca \PP^+\Bigr] 
\nonumber\\
\label{ll1}
&+&
\frac{h^{01}}{h^{00}}(\PP^+\x'^- + \PP^-\x'^+ +  \PP_\sca \y'^\sca)
-\frac{1}{h^{00}}(\PP^+ +h^{01}\x'^+)A^0
+\x'^+ A^1 \ , 
\\
\label{ll2}
{\cal L}_2
&=& \PP_\al\dot{x}_\al
+\frac{1}{2h^{00}}((\PP-C^0)_\al^2+e^{4\phi}\x'_\al^2)
+\frac{h^{01}}{h^{00}}(\PP-C^0)_\al\x'_\al +\x'_\al C_\al^1 \ , 
\\
\label{ll3}
{\cal L}_3
&= &
\Pi\dot{\phi} +\frac{1}{2h^{00}}e^{2\phi}(\Pi^2
+\ph'^2)
+\frac{h^{01}}{h^{00}}\Pi\ph'+ T \ . 
\end{eqnarray}
Next, we  impose the light-cone gauge

\be\label{lcg}
x^+ = \tau \,,\ \ \
\qquad
\PP^+=p^+ \ . 
\ee
Using  these  gauge conditions  in the 
action and integrating over $\PP^-$
we get the expression for  $h^{00}$

\be\label{h00}
h^{00} = -p^+ \  . 
\ee
Inserting this into   (\ref{ll1}), (\ref{ll2}) (\ref{ll3}) 
we get the following general  form of the 
 phase space \lc Lagrangian 

\begin{eqnarray}
{\cal L}_1
&=& 
\PP_\sca \dot{y}^\sca
-\frac{1}{2p^+}\Bigl(g^{\sca\scb}\PP_\sca \PP_\scb
+e^{4\phi}g_{\sca\scb}\y'^\sca \y'^\scb
+p^{+2}B - 2p^+ F^\sca \PP_\sca\Bigr)
\nonumber
\\
\label{genph1}
&-&
\frac{h^{01}}{p^+}(p^+\x'^-  +  \PP_\sca \y'^\sca)
+A^0 \ , 
\\
\label{genph2}
{\cal L}_2
&=& 
\PP_\al\dot{x}_\al - \frac{1}{2p^+}(\PP_\al^2+e^{4\phi}\x'_\al^2)
-\frac{h^{01}}{p^+}\PP_\al\x'_\al \ , 
\\
\label{genph3}
{\cal L}_3
&= &\Pi\dot{\phi} -\frac{1}{2p^+}e^{2\phi}(\Pi^2
+\ph'^2)
-\frac{h^{01}}{p^+}\Pi\ph' \ . 
\end{eqnarray}
 In deriving these  expressions we used the fact the functions 
$C^{\vm \al}$, $T$ given in (\ref{funC}),(\ref{funT}) are equal to zero 
in the light-cone gauge \rf{lcg}.
 This  general form of the phase space  Lagrangian can be
used  to derive  explicit forms of Lagrangians
 corresponding to different choices of bosonic
coordinates:
 one needs only to insert the appropriate functions $A^0$, $B$,
and $F^\sca$.
  For the ``intermediate" 
 case \rf{actkin3},\rf{actwz3}  these functions are given by
(\ref{funA}),(\ref{funB}),(\ref{funF}) so that we get 

\begin{eqnarray}
{\cal L} 
&= &
\PP_\al\dot{x}_\al +  \Pi\dot{\phi} 
+ \PP_\sca \dot{y}^\sca
+\frac{{\rm i}}{2}p^+(\theta^i \dot{\theta}_i
+\eta^i\dot{\eta}_i+\theta_i \dot{\theta}^i+\eta_i\dot{\eta}^i)
\nonumber\\
&-&
\frac{1}{2p^+}\Bigl[\PP_\al^2+e^{4\phi}\x'_\al^2
+e^{2\phi}(\Pi^2+\ph'^2
+l^i{}_j^{\,\,2} + G_{\sca\scb}\y'^\sca \y'^\scb
+p^{+2}(\eta^2)^2 + 4p^+\eta_i l^i{}_j \eta^j) 
\Bigr]
\nonumber\\
&+&
e^{2\phi}\eta^i C_{ij}^U
(\th'^j - {\rm i}\sqrt{2}e^\phi \eta^j\x')
+e^{2\phi}\eta_i C^{ij}_U
(\th'_j +{\rm i}\sqrt{2}e^\phi \eta_j\xb')
\nonumber\\
&-&
\frac{h^{01}}{p^+}\Bigl[p^+\x'^-  + \PP_\al\x'_\al + \Pi\ph'
+ \PP_\sca \y'^\sca
+\frac{{\rm i}}{2}p^+(\theta^i\th'_i+\eta^i\et'_i
+\theta_i\th'^i+\eta_i\et'^i)\Bigr] \   .
\end{eqnarray}
Here $C_U^{ij}= -(C_{ij}^U)^* $,  and  we  introduced the  notation

\be
l^i{}_j \equiv {\rm i}(V^\sca)^i{}_j \PP_\sca\ , 
\ee
and used  the relation

\be
G^{\sca\scb}\PP_\sca \PP_\scb =l^i{}_j^{\,\,2}\,,
\qquad
l^i{}_j^{\,\,2} \equiv l^i{}_j l^j{}_i\,.
\ee
By applying the coordinate transformation 
the above Lagrangian can be rewritten 
in the  form  corresponding to the case 
\rf{actkin6},\rf{actwz6} in which 
the $S^5$ part is parametrized  by the unit 6-d  vector 
$u^M$

\begin{eqnarray}\label{lag6}
{\cal L} 
&= &
\PP_\al\dot{x}_\al +  \Pi\dot{\phi} 
+ \PP_M \dot{u}^M
+\frac{{\rm i}}{2}p^+(\theta^i \dot{\theta}_i
+\eta^i\dot{\eta}_i+\theta_i
\dot{\theta}^i+\eta_i\dot{\eta}^i)
\nonumber\\
&-&
\frac{1}{2p^+}\Bigl[\PP_\al^2+e^{4\phi}\x'_\al^2
+e^{2\phi}(\Pi^2+\ph'^2
+l^i{}_j^{\,\,2} + \u'^M\u'^M
+p^{+2}(\eta^2)^2 + 4p^+\eta_i l^i{}_j \eta^j) 
\Bigr]
\nonumber\\
&+&
e^{2\phi}\eta^i y_{ij}
(\th'^j - {\rm i}\sqrt{2}e^\phi \eta^j\x')
+e^{2\phi}\eta_i y^{ij}
(\th'_j + {\rm i}\sqrt{2}e^\phi \eta_j\xb')
\nonumber\\
&-&
\frac{h^{01}}{p^+}\Bigl[p^+\x'^-  + \PP_\al\x'_\al + \Pi\ph'
+ \PP_M \u'^M
+\frac{{\rm i}}{2}p^+(\theta^i\th'_i+\eta^i\et'_i
+\theta_i\th'^i+\eta_i\et'^i)\Bigr] \  , 
\end{eqnarray}
where $\PP_M$ is the canonical momentum for $u^M$ and  (see Appendix B)

\be y_{ij}\equiv  \rho^M_{ij} u^M =C_{ij}^U\,,
\quad \ \ \ \  
y^{ij}\equiv  (\rho^M)^{ij} u^M =C^{ij}_U\,,
\ee \be 
l^i{}_j = { {\rm i}  \over 2} (\rho^{MN})^i{}_j u^M \PP^N \ .
\ee
Taking into account the constraint  $u^M \PP^M =0$  (see  (\ref{seccon}))
we get $l^i{}_kl^k{}_j=\frac{1}{4}\PP^M\PP^M\delta^i{}_j$.
The above Lagrangian gives the Hamiltonian 

\be\label{ham} H =-P^- \ , \ \ \ \ \ 
P^- =\int d \sigma\   \PP^- \ , 
\ee
where the Hamiltonian density $\PP^-$ is 

\begin{eqnarray}\label{strham}
\PP^-
&= &
-\frac{1}{2p^+}\Bigl[\PP_\al^2+e^{4\phi}\x'_\al^2
+e^{2\phi}(\Pi^2+\ph'^2
+l^i{}_j^{\,\,2} + \u'^M \u'^M
+p^{+2}(\eta^2)^2 + 4p^+\eta_i l^i{}_j \eta^j) 
\Bigr]
\nonumber\\
&+&
e^{2\phi}\eta^i y_{ij}
(\th'^j - {\rm i}\sqrt{2}e^\phi \eta^j\x')
\    +    \  e^{2\phi}\eta_i y^{ij}
(\th'_j + {\rm i}\sqrt{2}e^\phi \eta_j\xb')\, .
\end{eqnarray}
It should be supplemented by the  constraint

\be 
p^+\x'^-  + \PP_\al\x'_\al + \Pi\ph' + \PP_M \u'^M 
+\frac{{\rm i}}{2}p^+(\theta^i\th'_i+\eta^i\et'_i 
+\theta_i\th'^i+\eta_i\et'^i) =0\,.
\ee 
As usual,  this constraint  allows one  to express the non-zero modes of the 
bosonic
 coordinate $x^-$ in terms of the transverse physical  ones.

%We are now ready to compare this  Hamiltonian with the  superparticle one.  
It is easy
to see that in the particle theory limit the 
 superstring Hamiltonian  (\ref{strham})
reduces to the superparticle one  in (\ref{parham}).
The latter was found in
\ci{met3} by applying  the direct method of constructing relativistic dynamics
 \ci{dir} based on the symmetry algebra.\footnote{Strictly speaking, the  string Hamiltonian
(\ref{strham}) reduces to (\ref{parham})  modulo terms proportional to
$\eta^2$ and some constant. This difference is related to the fact that
here we are considering the   classical string Hamiltonian, 
i.e. ignore operator
ordering, while  the particle Hamiltonian (\ref{parham}) is 
defined  in
terms of quantum  operators.} 
The present discussion thus   provides a self-contained
derivation of the light-cone gauge superparticle 
 action from the covariant one.

%%%%%%%%%%%%%%%%%%%%%%%%%%%%%%%%%%%%%
\subsection{Equations of motion}
%%%%%%%%%%%%%%%%%%%%%%%%

The
equations of motion corresponding to  the phase space 
superstring Lagrangian \rf{lag6} take the following
form 
\begin{eqnarray}
\label{eqmot1}
\dot{x} 
&=&
\frac{1}{p^+}\PP\,,\ \ 
\qquad
\dot{\bar{x}} 
=
\frac{1}{p^+}\bar{\PP}\ ,  \ \ \ \ \ \ \ \   \dot{\phi} 
=
\frac{e^{2\phi}}{p^+}\Pi \ , 
\\
\dot{\PP}
&=&
\frac{1}{p^+}\partial_\sigma(e^{4\phi}\x')
-{\rm i}\sqrt{2}\partial_\sigma(e^{3\phi}\eta_i y^{ij}\eta_j)
\ , \\ 
\dot{\bar{\PP}}
&=&
\frac{1}{p^+}\partial_\sigma(e^{4\phi}\xb')
+{\rm i}\sqrt{2}\partial_\sigma(e^{3\phi}\eta^iy_{ij}\eta^j) \ , 
\\
\dot{\Pi}
&=& \frac{1}{p^+}\partial_\sigma (e^{2\phi}\ph')
-\frac{2}{p^+}e^{4\phi}\x'_\al^2
-\frac{e^{2\phi}}{p^+}\Bigl(\Pi^2+ \ph'^2+l^i{}_j^{\,\,2} 
+ \u'^M\u'^M
+p^{+2}(\eta^2)^2 + 4p^+\eta_i l^i{}_j \eta^j
\Bigr)
\nonumber\\
&+&
e^{2\phi}\eta^i y_{ij}
(2\th'^j - 3{\rm i}\sqrt{2}e^\phi \eta^j\x')
+e^{2\phi}\eta_i y^{ij}
(2\th'_j + 3{\rm i}\sqrt{2}e^\phi \eta_j\xb')\,,
\\
\dot{u}^M 
&=&\frac{e^{2\phi}}{p^+}\PP^M
-{\rm i}e^{2\phi}\eta_i(\rho^{MN})^i{}_j\eta^j u^N \ , 
\\
\dot{\PP}^M
& =& -\frac{e^{2\phi}}{p^+}u^M \PP^N\PP^N
+\frac{1}{p^+}v^{MN}\partial_\sigma(e^{2\phi}\u'^N)
-{\rm i}e^{2\phi}\eta_i(\rho^{MN})^i{}_j\eta^j \PP^N
\\
&+& 
e^{2\phi}v^{MN}\eta^i\rho_{ij}^N
(\th'^j-{\rm i}\sqrt{2}e^\phi \eta^j \x')
+e^{2\phi}v^{MN}\eta_i(\rho^N)^{ij}
(\th'_j+{\rm i}\sqrt{2}e^\phi \eta_j \xb')
\\
\dot{\theta^i}
&=&
-\frac{\rm i}{p^+}\partial_\sigma(e^{2\phi}y^{ij}\eta_j)
\ , \ \ \ \ \ \ 
\dot{\theta_i}
=
-\frac{\rm i}{p^+}\partial_\sigma(e^{2\phi}y_{ij}\eta^j) \ , 
\\
\dot{\eta}^i
&=&
e^{2\phi}
\Bigl[{\rm i}\eta^2\eta^i
-\frac{2\rm i}{p^+}(l\eta)^i
+\frac{\rm i}{p^+}
y^{ij}(\th'_j + {\rm i}2\sqrt{2}e^\phi\eta_j\xb')\Bigr]\ , 
\\
\label{eqmot11}
\dot{\eta}_i
&=&
e^{2\phi}
\Bigl[-{\rm i}\eta^2\eta_i
+\frac{2\rm i}{p^+}(\eta l)_i
+\frac{\rm i}{p^+}y_{ij}(\th'^j-{\rm i}2\sqrt{2}e^\phi\eta^j\x')\Bigr]\ . 
\end{eqnarray}
Here we defined 

\be
v^{MN} \equiv \delta^{MN} -u^M u^N\,,
\ee
and, as  previously,  do not distinguish between  the upper and lower 
indices $M,N$, i.e.  use  the convention $\PP_M = \PP^M$.
These equations can be written in the  Hamiltonian form.  
Introducing  
the notation $\cal X$
for the phase space variables  $(\PP_\al,x_\al,\Pi,\phi,\PP^M,u^M,
\theta^i,\theta_i,\eta^i,\eta_i)$ one has the Hamiltonian
equations

\be\label{hamfor}
\dot{\cal X} = [{\cal X},\PP^-] \ , 
\ee
where  the phase space variables
satisfy the (classical) Poisson-Dirac brackets

\be\label{qcom1}
[\ \PP(\sigma), \bar{x}(\sigma')\ ] =\delta(\sigma,\sigma')\,,
\quad
[\ \bar{\PP}(\sigma), x(\sigma')\ ] =\delta(\sigma,\sigma')\,,
\quad
[\ \Pi(\sigma), \phi(\sigma')\ ] =\delta(\sigma,\sigma')\,,
\ee
\be\label{npcom}
[\PP^M(\sigma),u^N(\sigma')]
=v^{MN}\delta(\sigma,\sigma')\,,
\qquad
[\PP^M(\sigma),\PP^N(\sigma')]
=(u^M \PP^N - u^N \PP^M ) \delta(\sigma,\sigma')\,,
\ee
%\be
%[\PP_{ij}(\sigma),y^{kl}(\sigma')]
%=\Bigl(2(\delta_i^l\delta_j^k-\delta_i^k\delta_j^l)
%-y_{ij}y^{kl}\Bigr)\delta(\sigma,\sigma') \ , 
%\ee
\be\label{ttcom}
\{\theta_i(\sigma), \theta^j(\sigma')\} 
= \frac{\rm i}{p^+}\delta_i^j\delta(\sigma,\sigma')\ , 
\qquad
\{\eta_i(\sigma), \eta^j(\sigma')\}  
=\frac{\rm i}{p^+}\delta_i^j\delta(\sigma,\sigma') \ , 
\ee
\be\label{qcom6}
[x_0^-,\theta^i]=\frac{1}{2p^+}\theta^i,
\quad
[x_0^-,\theta_i]=\frac{1}{2p^+}\theta_i\,,
\quad
[x_0^-,\eta^i]=\frac{1}{2p^+}\eta^i,
\quad
[x_0^-,\eta_i]=\frac{1}{2p^+}\eta_i\,,
\ee
where $x_0^-$ is the  zero mode of $x^-$. 
All  the remaining brackets are equal
to zero (with exception of $[p^+,x_0^-]=1$). 
The structure of  (\ref{npcom}) reflects the fact that in the Hamiltonian
formulation the constraint  $u^Mu^M=1$ should be supplemented by 
the constraint

\be\label{seccon}
u^M\PP^M=0\,.
\ee 
These  are second class constraints and the Dirac
procedure leads then to the classical Poisson-Dirac brackets (\ref{npcom}).
To derive     (\ref{ttcom}),(\ref{qcom6})  one is to take into account
 that  the Lagrangian 
(\ref{lag6}) has the following second class constraints

\be
 p_{\theta^i}+\frac{\rm i}{2}p^+\theta_i=0\,,
\qquad
 p_{\theta_i}+\frac{\rm i}{2}p^+\theta^i=0 \ , 
\ee
where $p_{\theta^i}$, $p_{\theta_i}$ are the canonical momenta of fermionic
coordinates. The same constraints are found for the fermionic coordinates $\eta^i$,
$\eta_i$.  Starting with the  Poisson brackets

\be
\{ p_{\theta^i},\theta^j\}_{_{P.B.}}=\delta_i^j\,,
\qquad
\{ p_{\theta_i},\theta_j\}_{_{P.B.}}=\delta^i_j\,,
\qquad
[p^+,x_0^-]_{_{P.B.}}=1\,,
\ee
one  gets then the 
 Poisson-Dirac brackets given in (\ref{ttcom}),(\ref{qcom6}).

%%%%%%%%%%%%%%%%%%%%%%%%%%%%%%%%%%%%%
\subsection{Light-cone  Hamiltonian for superstring in $AdS_3\times
S^3$  }
%%%%%%%%%%%%%%%%%%%%%%%%

Finally, let us note that  our results for  the  $AdS_5\times S^5$ 
string can
be generalized in a rather straightforward
way to the case of  superstring in 
$AdS_3\times S^3$  space  with RR 3-form background
%%NEW%%
(for the corresponding covariant GS action see \ci{adstri}).
% To get the bosonic light-cone   gauge action for this case  one 
%could
%use the $\kappa$ symmetry light-cone  action \ci{MT3} and then
%apply the
%procedure developed in this paper. 
Our  light-cone gauge action
is  written in the form which allows a straightforward
generalization to the $AdS_3\times S^3$ case: one is just to do a
dimensional reduction.

%Let us discuss the $AdS_3\times S^3$ Lagrangian and Hamiltonian
%using 
%(\ref{lag6}) and (\ref{strham}). 
To get the Lagrangian and Hamiltonian
$\PP^-$ in $AdS_3\times S^3$   case 
 we are  to set $x_\al=0$, $\PP_\al=0$
in  (\ref{lag6}), (\ref{strham}). Now instead of $su(4)\sim so(6)$
we have
$so(4)$ which is decomposed into $su(2)$ and $\tilde{su}(2)$. 
The fermionic
variables $\eta$ and $\theta$ are now   transforming  in the 
fundamental
representation of $SU(2)$ and $\tilde{SU}(2)$ respectively
(i.e.  the indices $i,j$ now take values $1,2$). The
charge conjugation matrix $\Csp_{ij}$ is now   given by 
$\Csp= c \sigma_2$,\ $|c|=1$.  

The  $AdS_3\times S^3$ superstring 
Lagrangian  then takes  the form

\begin{eqnarray}\label{lag7}
{\cal L}
&= &
\Pi\dot{\phi} 
+ \PP_M \dot{u}^M
+\frac{{\rm i}}{2}p^+(\theta^i \dot{\theta}_i
+\eta^i\dot{\eta}_i+\theta_i
\dot{\theta}^i+\eta_i\dot{\eta}^i)
+\PP^-
\\
&-&
\frac{h^{01}}{p^+}\Bigl[p^+\x'^-  +  \Pi\ph'
+ \PP_M \u'^M
+\frac{{\rm i}}{2}p^+(\theta^i\th'_i+\eta^i\et'_i
+\theta_i\th'^i+\eta_i\et'^i)\Bigr]\,.
\end{eqnarray}
The  (minus)  Hamiltonian density is given by
(cf. \rf{addss})

%\begin{eqnarray}
\be\label{strham3}
\PP^-
=-\frac{e^{2\phi}}{2p^+} 
\Bigl[
\Pi^2+\ph'^2
+2l^i{}_j^{\,\,2} + \u'^M \u'^M
+p^{+2}(\eta^2)^2 + 4p^+\eta_i l^i{}_j \eta^j\Bigr]
%\nonumber
%\\
%&+&
+e^{2\phi}(\eta^i y_{ij}\th'^j+ h.c.) 
\, , 
%\end{eqnarray}
\ee
where  

\be
y_{ij} = C_{ik}^\prime (\sigma^M)^k{}_j u^M\,,
\qquad
\sigma^M =(\sigma^1,\sigma^2,\sigma^3,-{\rm i}{\rm I})\,.
\ee
and $ M,N=1,2,3,4$.
The expression for orbital part of $SU(2)$ generators 
 $l^i{}_j$
takes the form

\be
l^i{}_j =\frac{{\rm i}}{2}(\sigma^{MN})^i{}_j u^M \PP^N\,,\ \ \ \ 
\quad
(\sigma^{MN})^i{}_j=\frac{1}{2}(\sigma^M)^i{}_k(\bar{\sigma}^N)^k{}_j
-(M\leftrightarrow N)\,,
\ee
where
$\bar{\sigma}^M 
=(\sigma^1,\sigma^2,\sigma^3,{\rm i}{\rm I})$. Note that in  this 
case one
has the 
relation $2l^i{}_j^2 =\PP^M\PP^M$,  and this explains the 
factor of 2 in
front of
$l^i{}_j^2$ in (\ref{strham3}).
The variation over $h^{01}$ gives the  constraint

\be 
p^+\x'^-  +  \Pi\ph'
+ \PP_M \u'^M
+\frac{{\rm i}}{2}p^+(\theta^i\th'_i+\eta^i\et'_i
+\theta_i\th'^i+\eta_i\et'^i)=0\,.
\ee
An  interesting feature of the  Hamiltonian \rf{strham3} 
 is that  $e^{2 \p}$ factors out  (cf. \rf{strham}). 
 Note also that  the dependence on matrix $C_{ij}^\prime$ 
can be  eliminated by   the   redefinition of the  fermionic
coordinates $\eta_i\rightarrow C_{ij}^\prime\eta^j$.

%%%%%%%%%%%%%%%%%%%%%%%%%%%%%%%%%%%%%%%%%%%%%%%%%%%%%%%
\newsection{Noether charges as generators of  supersymmetry algebra
 $psu(2,2|4)$ }
%%%%%%%%%%%%%%%%%%%%%%%%%%%%%%%%%%%%%%%%%%%%%%%%%%%%%%%%

The Noether charges play an  important role in 
the analysis of the symmetries of
dynamical systems. The choice of the light-cone gauge spoils  manifest global
symmetries,  and  in order to demonstrate that these  global invariances 
are still
present   one  needs to  find  the  Noether charges
which generate them.\foot{In what follows ``currents'' and ``charges''
will mean both bosonic and fermionic ones, 
i.e.  will include supercurrents  and supercharges.}
These charges
play  a crucial  role  in  formulating  superstring   field theory 
 in the \lc gauge  (see \ci{GSB,GS1}).\foot{Let us note  in passing that  the development  
of the 
light-cone string field theory
approach  in  the  case of the \adss  background 
 may be useful  in the context of AdS/CFT correspondence.
 One striking feature of the 
light-cone   closed superstring field theory actions  in the flat space case 
 is that their interaction part  contains only  
 cubic vertices.  Same may happen to be true  also in the  case of the 
  $AdS_5
\times S^5$ background, and that may  have important implications 
for  establishing correspondence with the 
SYM theory in the \lc gauge framework.}

The Noether charges for  a superparticle in $AdS_5\times S^5$ were  found
in \ci{met3}. These charges are helpful in
establishing 
%precise and manifestly  supersymmetric 
a correspondence
between the bulk fields of type  IIB supergravity and  
the chiral primary operators of the boundary  theory
in a manifestly  supersymmetric  way. 
Superstring Noether charges  should thus   be useful for the  study
 of the AdS/CFT correspondence at the full string-theory level.

In  the light-cone formalism the generators (charges)  of the  basic 
$psu(2,2|4)$ superalgebra  can be split into two groups:
\begin{equation}\label{kingen}
P^+,\,P,\,\bar{P},\, J^{+x},\,J^{+\bar x},\, K^+,\, K,\,{\bar K},\,
Q^{+i},\,Q^+_i,\,S^{+i},\,S^+_i,\,
D,\,J^{+-},\,J^{x\bar x}\,,
\end{equation}
which we  shall refer to as kinematic generators,  and
\begin{equation}\label{dyngen}
P^-,\, J^{-x}\,, J^{-\bar x},\,K^-\,, Q^{-i},\, Q^-_i,\, S^{-i},\,S^-_i \ , 
\end{equation}
which we shall  refer to as dynamical generators (see also \ci{MT3}).
The kinematic generators
have positive or zero $J^{+-}$ charges, while dynamical generators have
negative $J^{+-}$ charges.
 For $x^+=0$ the kinematic generators in
the superfield realization are
quadratic in the  physical    string fields
(i.e. they have the structure    $ J= J_1 + x^+J_2+x^{+2}J_3$ where     
$J_1$  is quadratic  but $J_2$, $J_3$ contain higher order terms 
  in second-quantized fields), while the dynamical
generators receive higher-order interaction-dependent  corrections.  
The first step in  the  construction  of superstring field 
theory is to find  a  free (quadratic) superfield 
 representation of the generators of the $psu(2,2|4)$
superalgebra.
The   charges we  obtain  below
can be used  to  obtain  (after quantization) 
     these free superstring field  charges.

%%%%%%%%%%%%%%%%%%%%%%%%
\subsection{Currents for  $\kappa$-symmetry 
light-cone  gauge  fixed  superstring action}
%%%%%%%%%%%%%%%%%%%%%%%%%%%

As usual,   symmetry  generating charges   can be obtained from conserved currents. 
Since currents  themselves 
may be helpful in  some  applications, we shall  first
derive  them starting with the  
$\kappa$-symmetry gauge fixed  Lagrangian in the form given in 
(\ref{actkin6}),(\ref{actwz6}). To obtain the currents we shall  use the 
standard Noether method (see, e.g., \ci{W})  based on the localization of 
the parameters of
the  associated global transformations.
 Let $\epsilon$ be a parameter of some global transformation 
which leaves the action  invariant. Replacing it by a 
function of worldsheet coordinates $\tau,\sigma$,   the variation of  the 
action takes the form

\be\label{dS}
\delta S =\int d^2\sigma\  {\cal G}^\vm \partial_\vm \epsilon\, , 
\ee
where ${\cal G}^\vm$  is the  corresponding current.
Making use of this formula, we shall  find below those currents which
are related to  symmetries that do not 
involve compensating $\kappa$-symmetry transformation. 
The remaining currents will  be found  in the next subsection 
starting  from  the 
action   \rf{lag6}  where   both the   $\kappa$-symmetry and 
the bosonic  light-cone gauges are fixed.

Let us  start with  the translation invariance 
$\delta x^a = \epsilon^a$.
Applying  (\ref{dS}) to the Lagrangian 
(\ref{actkin6}),(\ref{actwz6}) gives the translation currents
\begin{eqnarray}
\label{con1}
{\cal P}^{+\vm}
&=&-\sqrt{g}g^{\vm\vn}e^{2\phi}\partial_\vn x^+\ , 
\\
{\cal P}^\vm 
&=&
-\sqrt{g}g^{\vm\vn}e^{2\phi}\partial_\vn x
-{\rm i}\sqrt{2}e^{3\phi}\epsilon^{\vm\vn}\eta_i y^{ij}\eta_j
\partial_\vn x^+\ , 
\\
\label{con3}
\bar{{\cal P}}^\vm 
&=&-\sqrt{g}g^{\vm\vn}e^{2\phi}\partial_\vn \bar{x}
+{\rm i}\sqrt{2}e^{3\phi}\epsilon^{\vm\vn}\eta^i y_{ij}\eta^j
\partial_\vn x^+\ , 
\\
{\cal P}^{-\vm}
&=&-\sqrt{g}g^{\vm\vn}\Bigl(e^{2\phi}\partial_\vn x^-
+F^M D_\vn u^M\Bigr) \nonumber
\\
&-&
\frac{\rm i}{2}\sqrt{g}g^{\vm\vn}e^{2\phi}\Bigr(
\theta^i\partial_\vn\theta_i
+\theta_i\partial_\vn\theta^i
+\eta^i\partial_\vn\eta_i
+\eta^i\partial_\vn\eta^i+2{\rm i}e^{2\phi}\partial_\vn x^+(\eta^2)^2\Bigr)
\nonumber
\\
&+&
\epsilon^{\vm\vn}e^{2\phi}\eta^i y_{ij}(\partial_\vn\theta^j
-{\rm i}\sqrt{2}e^\phi\eta^j\partial_\vn x)
+\epsilon^{\vm\vn}e^{2\phi}\eta_i y^{ij}(\partial_\vn\theta_j
+{\rm i}\sqrt{2}e^\phi\eta_j\partial_\vn \bar{x})\,,
\label{con4}
\hspace{1cm}
\end{eqnarray}
where
\be
F^M \equiv {\rm i}\eta_i(\rho^{MN})^i{}_j \eta^j e^{2\phi}u^N\,.
\ee
Invariance of the action (\ref{actkin6}),(\ref{actwz6}) 
with respect to rotations in $(x^-,x)$ and
$(x^-,\bar{x})$ planes

\be
\delta \bar{x} = \epsilon_{_J} x^+\,,
\quad
\delta x^- =- \epsilon_{_J} x\,,
\qquad
\delta x = \epsilon_{_{\bar{J}}} x^+\,,
\quad
\delta x^- =- \epsilon_{_{\bar{J}}} \bar{x}\,,
\ee
gives the following conserved currents

\be\label{con5}
{\cal J}^{+x} = -\sqrt{g}g^{\vm\vn}e^{2\phi}(x^+\partial_\vn x 
-x \partial_\vn x^+)
-{\rm i}x^+\epsilon^{\vm\vn}e^{3\phi}\partial_\vn x^+ \eta_i y^{ij} \eta_j \ , 
\ee
\be\label{con6}
{\cal J}^{+\bar{x}} = -\sqrt{g}g^{\vm\vn}e^{2\phi}(x^+\partial_\vn \bar{x} 
-\bar{x} \partial_\vn x^+)
+{\rm i}x^+\epsilon^{\vm\vn}e^{3\phi}\partial_\vn x^+ \eta^i y_{ij} \eta^j\,.
\ee
Making use  of  (\ref{con1})-(\ref{con3})  we get 

\be\label{con7}
{\cal J}^{+x\vm}=  x^+{\cal P}^\vm - x{\cal P}^{+\vm}\,,
\qquad
{\cal J}^{+\bar{x}\vm}=  x^+\bar{{\cal P}}^\vm - \bar{x}{\cal P}^{+\vm} \ . 
\ee
Some of the  remaining bosonic currents can be expressed 
 in terms of supercurrents.
The invariance with respect to the  super-transformations

\be
\delta\theta^i = \epsilon^i\,,
\qquad
\delta\theta_i  = \epsilon_i\,,
\qquad
\delta x^- 
= -\frac{\rm i}{2}\epsilon^i\theta_i -\frac{\rm i}{2}\epsilon_i\theta^i \ , 
\ee
gives the following supercurrents

\be\label{con8}
{\cal Q}^{+i\vm}=-\sqrt{g}g^{\vm\vn}e^{2\phi}\theta^i\partial_\vn x^+
-{\rm i}\epsilon^{\vm\vn}e^{2\phi}y^{ij}\eta_j\partial_\vn x^+ \ , 
\ee
\be\label{con9}
{\cal Q}_i^{+\vm}=-\sqrt{g}g^{\vm\vn}e^{2\phi}\theta_i\partial_\vn x^+
-{\rm i}\epsilon^{\vm\vn}e^{2\phi}y_{ij}\eta^j\partial_\vn x^+ \ . 
\ee
%Now we return to remaining bosonic currents.
The invariance of the action (\ref{actkin6}),(\ref{actwz6}) 
with respect to the rotation of (super)coordinates 
in the  $(x^+,x^-)$ plane

\be
\delta x^\pm =e^{\pm \epsilon} x^\pm\,,
\qquad
\delta(\theta^i,\theta_i\,,\eta^i,\eta_i)
=e^{-\epsilon/2}(\theta^i,\theta_i\,,\eta^i,\eta_i) \ , 
\ee
gives the following conserved current

\begin{eqnarray}\label{con10}
{\cal J}^{+-\vm}
&=&
-\sqrt{g}g^{\vm\vn}e^{2\phi}(x^+\partial_\vn x^-
-x^-\partial_\vn x^+)
\nonumber\\
&-&
\frac{\rm i}{2}\sqrt{g}g^{\vm\vn}e^{2\phi}x^+
\Bigl(\theta^i\partial_\vn\theta_i
+\theta_i\partial_\vn\theta^i
+\eta^i\partial_\vn\eta_i
+\eta^i\partial_\vn\eta^i+2{\rm i}e^{2\phi}\partial_\vn x^+(\eta^2)^2\Bigr)
\nonumber\\
&+&
x^+\epsilon^{\vm\vn}e^{2\phi}\eta^i y_{ij}(\partial_\vn\theta^j
-{\rm i}\sqrt{2}e^\phi\eta^j\partial_\vn x)
+x^+\epsilon^{\vm\vn}e^{2\phi}\eta_i y^{ij}(\partial_\vn\theta_j
+{\rm i}\sqrt{2}e^\phi\eta_j\partial_\vn \bar{x})
\nonumber\\
&-&
\frac{1}{2}\epsilon^{\vm\vn}e^{2\phi}\partial_\vn x^+ \eta^i y_{ij}\theta^j
-\frac{1}{2}
\epsilon^{\vm\vn}e^{2\phi}\partial_\vn x^+ \eta_i y^{ij}\theta_j \ . 
\end{eqnarray}
This current can be represented  in terms of  the translation currents
as follows

\be\label{con11}
{\cal J}^{+-\vm}
=x^+{\cal P}^{-\vm}- x^-{\cal P}^{+\vm}
+\frac{\rm i}{2}\theta^i{\cal Q}_i^{+\vm}
+\frac{\rm i}{2}\theta_i{\cal Q}^{+i\vm} \ . 
\ee
The invariance with respect  to the  rotation of (super)coordinates in 
$(x,\bar{x})$  plane 

$$
\delta x=e^{{\rm i}\epsilon}x\,,
\quad
\delta \bar{x}=e^{-{\rm i}\epsilon}x\,,
\quad
\delta \theta^i=e^{\frac{\rm i\epsilon}{2}}\theta^i\,,
\quad
\delta \theta_i=e^{-\frac{\rm i\epsilon}{2}}\theta_i\,,
\quad
\delta \eta^i=e^{-\frac{\rm i\epsilon}{2}}\eta^i\,,
\quad
\delta \eta_i=e^{\frac{\rm i\epsilon}{2}}\eta_i\,,
$$
leads to the following conserved current

\begin{eqnarray}
{\cal J}^{x\bar{x}\vm}
&=& -\sqrt{g}g^{\vm\vn}e^{2\phi}
(x\partial_\vn \bar{x} - \bar{x}\partial_\vn x)
+\frac{\rm i}{2}\sqrt{g}g^{\vm\vn}\partial_\vn x^+
(\theta^i\theta_i-\eta^i\eta_i)
\nonumber
\\
&-&\frac{1}{2}\epsilon^{\vm\vn}
e^{2\phi}\partial_\vn x^+ \eta^i y_{ij}(\theta^j-{\rm i}2\sqrt{2}\eta^jx)
+\frac{1}{2}\epsilon^{\vm\vn}
e^{2\phi}\partial_\vn x^+ \eta_i y^{ij}
(\theta_j+{\rm i}2\sqrt{2}\eta_j\bar{x}) \ , 
\label{con12}
\end{eqnarray}
which can be rewritten  in terms  of translation 
currents as follows

\be\label{con13}
{\cal J}^{x\bar{x}\vm}
=x\bar{\PP}^\vm - \bar{x}\PP^\vm 
-\frac{\rm i}{2}\theta^i{\cal Q}_i^{+\vm}
+\frac{\rm i}{2}\theta_i{\cal Q}^{+i\vm}
+\frac{\rm i}{2}(\theta^i\theta_i+\eta^i\eta_i)\PP^{+\vm}
\ee
The invariance with respect to  the dilatations

\be
\delta x^a = e^\epsilon x^a\,,
\qquad
\delta\phi =-\epsilon\,,
\qquad
\delta(\theta^i,\theta_i\,,\eta^i,\eta_i)
=e^{\epsilon/2}(\theta^i,\theta_i\,,\eta^i,\eta_i) \ , 
\ee
leads to  the dilatation current

\begin{eqnarray}\label{con14}
{}\hspace{-1.5cm}{\cal D}^\vm
&=&
-\sqrt{g}g^{\vm\vn}\Bigl(e^{2\phi}x^a\partial_\vn x^a
+F^MD_\vn u^M -\partial_\vn\phi\Bigr)
\nonumber\\
&-&
\frac{\rm i}{2}\sqrt{g}g^{\vm\vn}e^{2\phi}x^+
\Bigl(\theta^i\partial_\vn\theta_i
+\theta_i\partial_\vn\theta^i
+\eta^i\partial_\vn\eta_i
+\eta^i\partial_\vn\eta^i+2{\rm i}e^{2\phi}\partial_\vn x^+(\eta^2)^2\Bigr)
\nonumber\\
&+&
x^+\epsilon^{\vm\vn}e^{2\phi}\eta^i y_{ij}(\partial_\vn\theta^j
-{\rm i}\sqrt{2}e^\phi\eta^j\partial_\vn x)
+x^+\epsilon^{\vm\vn}e^{2\phi}\eta_i y^{ij}(\partial_\vn\theta_j
+{\rm i}\sqrt{2}e^\phi\eta_j\partial_\vn \bar{x})
\nonumber\\
&-&
\frac{\epsilon^{\vm\vn}}{2}e^{2\phi}\partial_\vn x^+ \eta^i y_{ij}
(\theta^j-{\rm i}2\sqrt{2}e^\phi\eta^j x)
-\frac{\epsilon^{\vm\vn}}{2}e^{2\phi}\partial_\vn x^+ \eta_i y^{ij}
(\theta_j+{\rm i}2\sqrt{2}e^\phi\eta_j\bar{x})\,.
\hspace{1cm}
\end{eqnarray}
This current can be rewritten as

\be\label{con15}
{\cal D}^\vm
=x^a{\cal P}^{a\vm}+\sqrt{g}g^{\vm\vn}\partial_\vn \phi
-\frac{\rm i}{2}\theta^i{\cal Q}_i^{+\vm}
-\frac{\rm i}{2}\theta_i{\cal Q}^{+i\vm} \ . 
\ee
The invariance with respect to  the $SU(4)$ rotations
($ \epsilon^i{}_i=0$) 
\be   
\delta y^{ij} = \epsilon^i{}_l y^{lj}
+\epsilon^j{}_l y^{il} 
 \ , \ \ \ \ \ 
{\rm i.e.}  
 \ \ \ \  \delta u^M =-\frac{1}{2}\epsilon^i{}_j
  (\rho^{MN})^j{}_i u^N 
\ , 
\ee
\be
\delta \theta^i =\epsilon^i{}_j\theta^j\,,
\qquad \ \ \  
\delta \theta_i =-\theta_j \epsilon^j{}_i\,,
\qquad
\delta \eta^i =\epsilon^i{}_j\eta^j\,,
\qquad \ \ \  
\delta \eta_i =-\eta_j \epsilon^j{}_i\,,
\ee
gives the following $SU(4)$ current

\begin{eqnarray}\label{con16}
{}\hspace{-1cm}
{\cal J}^i{}_j^{ \ \vm  } 
=&-&{\rm i}\sqrt{g}g^{\vm\vn}\frac{1}{2}(\rho^{MN})^i{}_ju^MD_\vn u^N
+[\theta^i\theta_j+\eta^i\eta_j-\frac{1}{4}\delta^i_j(
\theta^l\theta_l+\eta^l\eta_l)]\PP^{+\vm}
\nonumber\\
&-&{\rm i}
\epsilon^{\vm\vn}e^{2\phi}\partial_\vn x^+(\theta^iy_{jl}\eta^l
-\frac{1}{4}\delta^i_j\theta^k y_{kl}\eta^l)
+{\rm i}\epsilon^{\vm\vn}e^{2\phi}\partial_\vn x^+(\theta_j y^{il}\eta_l
-\frac{1}{4}\delta^i_j \theta_k y^{kl}\eta_l)\,.
\end{eqnarray}

%%%%%%%%%%%%%%%%%%%%%%%%%%%%%%%%%
\subsection{Charges for  bosonic and $\kappa$-symmetry 
  light-cone gauge fixed 
superstring action}
%%%%%%%%%%%%%%%%%%%%%%%%%%%%%%%%%%

In  the previous section we have found (super)currents starting with the  $\kappa$-symmetry 
\lc gauge fixed 
action given in (\ref{actkin6}),(\ref{actwz6}). 
These currents can be used  to find currents for  the  action where 
both the fermionic $\kappa$-symmetry and the 
bosonic  reparametrization symmetry are fixed by the 
\lc type gauges   (\ref{lag6}). 
To find the components of currents in the world-sheet time direction 
${\cal G}^0$
one needs to use the relations  
(\ref{can1})--(\ref{can6})  for  the canonical momenta and to 
  insert the  light-cone gauge conditions  (\ref{lcg}) and (\ref{h00}) 
into   the expressions for the currents given in the previous subsection. 
The  charges are then given by

\be
G =\int d\sigma\  {\cal G}^0 \ . 
\ee
Let us start with the kinematic generators (charges) (\ref{kingen}). 
The results for the currents imply the following representations
for some of them

\be\la{kin1}
P = \int \PP\,,
\qquad
\bar{P} =  \int\bar{\PP}\,,
\qquad
P^+ = p^+\ , 
\ee
\be\la{kin2}
J^{+x } =\int x^+\PP -x p^+\,,
\qquad
J^{+\bar{x}} =\int x^+\bar{\PP} -\bar{x} p^+\,,
\ee
\be\la{kin3}
Q^{+i} = \int p^+ \theta^i\,,
\qquad
Q_i^+ = \int  p^+\theta_i\ . 
\ee
Note that these charges depend only on the zero modes of string
coordinates. 
In (\ref{kin1})--(\ref{kin3}) 
the integrands are ${\cal G}^0$ parts
of the corresponding currents in world-sheet time direction:
${\cal P}^0$,
$\bar{\cal P}^0$,
${\cal J}^{+x 0}$,
${\cal J}^{+\bar{x} 0}$,
${\cal Q}^{+i 0}$,
${\cal Q}_i^{+ 0}$
and
${\cal P}^{+0}=p^+$.
The components of currents in  the world-sheet 
spatial  direction ${\cal G}^1$ 
can be found simply by using the conservation laws,
the expressions for ${\cal G}^0$ 
and the  equations of
motion (\ref{eqmot1})--(\ref{eqmot11}). 
In this way we  obtain  

\be
{\cal P}^1 = -\frac{1}{p^+}e^{4\phi}\x'
+{\rm i}\sqrt{2}e^{3\phi}\eta_i y^{ij} \eta_j\,,
\qquad
\bar{{\cal P}}^1 = -\frac{1}{p^+}e^{4\phi}\xb'
-{\rm i}\sqrt{2}e^{3\phi}\eta^i y_{ij} \eta^j\ , 
\ee
\be
{\cal J}^{+x 1} = x^+\PP^1\,,
\qquad
{\cal J}^{+\bar{x} 1} = x^+\bar{\PP}^1\ , 
\qquad
{\cal Q}^{+i 1}= \frac{\rm i}{p^+}e^{2\phi}y^{ij}\eta_j\,,
\qquad
{\cal Q}_i^{+1}= \frac{\rm i}{p^+}e^{2\phi}y_{ij}\eta^j\,.
\ee
The remaining kinematic charges depend on non-zero string modes and
are given by

\begin{eqnarray}
\la{diag1}&&
J^{x\bar{x}} = \int x \bar{\PP} - \bar{x}\PP
-\frac{\rm i}{2}p^+\theta^2 +\frac{\rm i}{2}p^+\eta^2 \ , 
\\
\label{diag2}
&&
J^i{}_j = \int l^i{}_j +p^+\theta^i \theta_j + p^+\eta^i\eta_j
-\frac{1}{4}\delta_j^i p^+(\theta^2 + \eta^2)\ , 
\label{diag3}
\\
&&
J^{+-} = \int x^+\PP^- -x^- p^+ \ , 
\\
\la{diag4}
&&
D = 
\int 
x^+\PP^-  + x^- p^+  + \bar{x}\PP + \bar{x}\PP - \Pi\ . 
\end{eqnarray}
The derivation of the  remaining charges can be found in Appendix D.
The expressions for the  conformal supercharges are given by

\be 
S^+_i=S_i^+|_{x^+=0} -{\rm i}x^+ Q^-_i\,,
\qquad
S^{+i}=S^{+i}|_{x^+=0} + {\rm i}x^+ Q^{-i}\,,
\ee
where the ${x^+=0}$ parts (cf. (\ref{g0}))
are given by

\be\label{den1}
S_i^{+ }|_{x^+=0}
=
\int \frac{1}{\sqrt{2}}e^{-\phi} p^+\eta_i +{\rm i}p^+\theta_i x\,,
\qquad
S^{+i }|_{x^+=0}
= 
\int \frac{1}{\sqrt{2}}e^{-\phi} p^+\eta^i - {\rm i}p^+\theta^i \bar{x}\,.
\ee
The  Poincar\'e supercharges $Q^{-i}$, $Q^-_i$
are

\begin{eqnarray}
\label{q1}
&&
Q_i^- =\int \PP \theta_i 
+\frac{e^\phi}{\sqrt{2}}
\Bigl({\rm i}\eta_i\Pi
-p^+\eta^2\eta_i + 2 (\eta l)_i
+y_{ij}(\th'^j -{\rm i}\sqrt{2}e^\phi\eta^j\x')\Bigr) \ , 
\\
\label{q2}&&
Q^{-i} =\int
\bar{\PP}\theta^i
+\frac{e^\phi}{\sqrt{2}}
\Bigl(-{\rm i}\eta^i\Pi
-p^+\eta^2\eta^i + 2 (l\eta)^i
-y^{ij}(\th'_j +{\rm i}\sqrt{2}e^\phi\eta_j
\xb')\Bigr)\ , 
\end{eqnarray}
where

\be
(\eta l)_i \equiv \eta_j l^j{}_i\,,
\qquad \ \ \ \ 
(l\eta)^i \equiv l^i{}_j\eta^j\,.
\ee
The conformal boost charges can be represented as follows (see Appendix D)

\begin{equation}
K^+ =K^+|_{x^+=0} + x^+ (D + J^{+-})|_{x^+=0} + x^{+2}P^-\,,
\ee
\begin{equation}\label{adskmcom}
K=K|_{x^+=0} - x^+ J^{-x}\,,
\ee
where the parts  that  
do not depend  on $x^+$ are 

\begin{eqnarray}
&&
K^+|_{x^+=0} =
\int
-\frac{1}{2}(e^{-2\phi} + 2x\bar{x})p^+\,,
\\
&&
K|_{x^+=0}= 
\int 
-\frac{1}{2}e^{-2\phi}\PP
+ x (x^- p^+  + x\bar{\PP} - \Pi
+\frac{\rm i}{2}p^+\theta^2 +\frac{\rm i}{2}p^+\eta^2)
- \theta^i{\cal S}_i^{+0}|_{x^+=0}\,,\hspace{1cm}
\\
&&
\bar{K}|_{x^+=0}= 
\int 
-\frac{1}{2}e^{-2\phi}\bar{\PP}
+ \bar{x} (x^- p^+  + \bar{x}\PP  - \Pi
- \frac{\rm i}{2}p^+\theta^2 - \frac{\rm i}{2}p^+\eta^2)
+\theta_i{\cal S}^{+i 0}|_{x^+=0}\,.
\end{eqnarray}
Here  the densities of conformal supercharges ${\cal S}_i^{+0}|_{x^+=0}$,
${\cal S}^{+ i0}|_{x^+=0}$ are given by the integrands in 
(\ref{den1}).

Note that the $G|_{x^+=0}$ parts of the kinematic charges can be  obtained
from the  superparticle  ones simply by replacing the 
 particle coordinates by the
string ones.
The remaining dynamical generators  $J^{-x}$, $J^{-\bar{x}}$, 
$S^{-i}$, $S^-_i$, 
$K^-$ can be found  by using the expression found above
and  applying  the  commutation 
relations of $psu(2,2|4)$ superalgebra.
% Because these expressions are
%not illuminating we do not present them here.

The structure of the 
 $psu(2,2|4)$ generators  we have presented  in this section 
  is, of course, more complicated than  found 
in the  flat space case.
 But still  there are some interesting 
simplifications which have  an algebraic origin.
A  remarkable  feature of $psu(2,2|4)$ 
as compared to the Poincar\'e
superalgebra  is that in order to find
all  of its dynamical generators  it is sufficient to know only the 
kinematic ones  (\ref{kingen}) and  the Hamiltonian $P^-$ (\ref{ham}):
the  dynamical generators are obtained  using  commutation relations
between  the kinematic generators  and the 
Hamiltonian $P^-$.

%%%%%%%%%%%%%%%%%%%%%%%%%%%%%%%%%%%%%%%%%%%%%%%%%
\section*{Acknowledgments}
%%%%%%%%%%%%%%%%%%%%%%%%%%%%%%%%%%%%%%%%%%%%%%%%%%

The  work  of R.R.M. and A.A.T.  was  supported  by
the DOE grant DE-FG02-91ER-40690 
and the INTAS project 991590.
 R.R.M. is also supported by the RFBR Grant No.99-02-16207.
A.A.T. would like to acknowledge also the support of the EC TMR 
grant ERBFMRX-CT96-0045 and the PPARC SPG grant  PPA/G/S/1998/00613,
and acknowledges also the hospitality of the Aspen Center 
for Physics
and the  CERN Theory Division 
during the completion of this paper.
The work of  C.B.T. was supported by  the 
Department of Energy under Grant No. DE-FG02-97ER-41029.

\setcounter{section}{0}
\setcounter{subsection}{0}

%%%%%%%%%%%%%%%%%%%%%%%%%%%%%%%%%%%%%%%%%%%%%%%%

\appendix{Notation}
\label{not}

In the main part of the paper we 
 use the following  conventions for the indices:
\begin{eqnarray*}
a, b, c =0,\ldots, 3 & &\qquad
\hbox{boundary Minkowski space indices}
\\
{\cal A}, {\cal B}, {\cal C}=1,\ldots, 5 & &\qquad
\hbox{$S^5$  coordinate space indices}
\\
\apr,\bpr,\cpr=1,\ldots, 5 && \qquad  so(5)\  \hbox{ vector
indices ($S^5$ tangent space indices) }
\\
M,N,K,L =1,\ldots, 6 
&& \qquad so(6)\  \hbox{vector indices}
\\
i,j,k,l =1,\ldots, 4 && \qquad su(4)\  \hbox{vector indices}
\\
\mu,\nu = 0,1 
&&
\qquad
\hbox{world sheet coordinate indices}
\end{eqnarray*}
We decompose $x^a$ into the light-cone and  2 complex coordinates:
$x^a= (x^+,x^-,x, \bx)$

\be
x^\pm\equiv \frac{1}{\sqrt{2}}(x^3\pm x^0)\,,
\qquad \ \ \
x,\bx = { 1 \ov \sqrt 2} (x^1 \pm  {\rm i} x^2)\,.
%\qquad \bar x={ 1 \ov \sqrt 2} (x^1 - {\rm i} x^2)\,,
\end{equation}
We suppress the flat space metric tensor $\eta_{ab}=(-,+,+,+)$ 
in scalar products, i.e. 
$A^a B^a\equiv \eta_{ab}A^a B^b $.
The $SO(3,1)$ vector $A^a$  is decomposed as
$A^a = (A^+,A^-,A^x,A^{\bar{x}})$  so that the scalar product is 

\be
A^a B^a = A^+B^- + A^- B^+ +A\bar{B} +\bar{A} B \ ,
\ee
where we use the  convention

\be
A \equiv A^x =A_{\bar{x}}\,,
\qquad
\bar{A} \equiv A^{\bar{x}} = A_x\,.
\ee
We use the notation $x^\al$ to represent  $(x,\bar{x})$  with the 
following summation rule

\be
x^\al z^\al= x\bar{z}+\bar{x}z\,.
\ee
The  derivatives with respect to the  world-sheet coordinates
$(\tau,\sigma)$ are 

\be
\dot{x} \equiv \partial_\tau x\,,
\qquad
\x' \equiv \partial_\sigma x
\ee
The  world-sheet 
Levi-Civita $\epsilon^{\vm\vn}$   is defined with   $\epsilon^{01}=1$.

The six  matrices   $\rho_{ij}^M$  represent 
the $SO(6)$   Dirac   matrices $\gamma^M$
in the  chiral representation, i.e. 
\be\label{usgam}
\gamma^M
=\left(\begin{array}{cc}
 0   & (\rho^M)^{ij} 
 \\
 \rho_{ij}^M & 0
 \end{array}
 \right)\,, \ee
 \be 
 (\rho^M)^{il}\rho_{lj}^N + (\rho^N)^{il}\rho_{lj}^M
 =2\delta^{MN}\delta_j^i\,,
 \qquad
 \rho_{ij}^M =- \rho_{ji}^M\,,
 \qquad (\rho^M)^{ij}\equiv  - (\rho_{ij}^{M})^* \ . 
 \ee
The $SO(5)$  Dirac  and  charge conjugation 
 matrices  can be expressed  in terms 
of the $\rho^M$ matrices as   follows 

\be\label{Cg}
(\gamma^\apr)^i{}_j = {\rm i}(\rho^\apr)^{il}\rho_{lj}^6\,,\ \ 
\qquad
C_{ij}^\prime =\rho_{ij}^6 \  . 
\ee
The $\rho^M$ matrices satisfy the identities

\be
\rho_{ij}^M=\frac{1}{2}\epsilon_{ijkl}(\rho^M)^{kl}\,,
\qquad
\rho_{ij}^M(\rho^M)^{kl}
=2(\delta_i^l\delta_j^k-\delta_i^k\delta_j^l) \ . 
\la{iii}
\ee
The matrices $\rho^{MN}$  are defined by 

\be
(\rho^{MN})^i{}_j \equiv \frac{1}{2}(\rho^M)^{il}\rho_{lj}^N
-(M\leftrightarrow N)\,, 
\la{rrr}
\ee
so that 

\be
(\rho^{MN})^i{}_j (\rho^{MN})^k{}_l= 2\delta^i_j\delta^k_l
                                    -8\delta^i_l\delta^k_j\,.
\ee
We use the following hermitian conjugation rule for the 
fermionic coordinates

\be
\theta_i^\dagger =\theta^i\,,
\qquad
\eta_i^\dagger = \eta^i\ , 
\ee
and the following  notation  for their squares 

\be
\theta^2 \equiv \theta^i\theta_i\,,
\qquad \ \ \  
\eta^2 \equiv \eta^i\eta_i\,.
\ee

%%%%%%%%%%%%%%%%%%%%%%%%%%%%%%%%%%%%%%%%%
\appendix{Relation  between 
%forms of  string action
  different \\ parametrizations 
of $S^5$ }
%%%%%%%%%%%%%%%%%%%%%%%%%%%%%%%%

In the superstring Lagrangian in (\ref{actwz3}),(\ref{actwz3}) we
use the 5 independent 
$S^5$ coordinates $y^\sca$ in terms of which the 5-sphere  interval, 
metric tensor
and  vielbein are given by
\be
ds_{S^5}^2=d|y|^2+ {\rm sin}^2|y| ds_{S^4}^2 \ , \ \ \ \  
\    ds_{S^4}^2 = dn^\sca dn^\sca \   ,  \ \ \ n^\sca n^\sca =1 \ ,  
\ee 
\begin{equation}
G_{\sca\scb}= e_\sca^\apr e_\scb^\apr\,,   \ \
\qquad
e_\sca^\apr=\frac{\sin |y|}{|y|}(\delta_\sca^\apr-n_\sca n^\apr)
+n_\sca n^\apr    \ , 
\end{equation}
\be\label{b2}
G_{\sca\scb}= 
\frac{\sin^2|y|}{|y|^2}(\delta_{\sca\scb}-n_\sca n_\scb)
+n_\sca n_\scb \ ,    
\qquad
n^\sca \equiv \frac{y^\sca}{|y|}\,,
\qquad
|y| =\sqrt{y^\apr y^\apr}\,.
\ee
We use the convention
 $y^\sca = \delta_\apr^\sca y^\apr$ and the same
for $n^\apr$. The $S^5$  Killing vectors  $V^\apr$ and
$V^{\apr\bpr}$ corresponding to the five  translations and ten  
 $SO(5)$ rotations
respectively are  given  by 
\begin{eqnarray}
&&
V^\apr
=\Bigl[|y|\cot |y| (\delta^{\apr\sca}-n^\apr n^\sca)
+n^\apr n^\sca\Bigr]\partial_{y^\sca}  \ ,
\\
\label{b4}
&&
V^{\apr\bpr}=y^\apr \partial_{y^\bpr}-y^\bpr \partial_{y^\apr}  \ .
\end{eqnarray}
They can be collected into the  $SU(4)$ combination

\begin{equation}\label{b5}
(V^\sca)^i{}_j\partial_{y^\sca}
=\frac{1}{4}(\gamma^{\apr\bpr})^i{}_jV^{\apr\bpr}
+\frac{\rm i}{2}(\gamma^\apr)^i{}_j V^\apr      \ .
\end{equation}
The  commutation relations of $so(6)$ algebra  then include

\be
[V^\apr,V^\bpr]=-V^{\apr\bpr}\,,
\qquad
[V^\apr,V^{\bpr\cpr}]=\delta^{\apr\bpr}V^\cpr
-\delta^{\apr\cpr}V^\bpr \ . 
\ee
The matrix $C_{ij}^U$ in the  WZ part of the action 
(\ref{actwz3}) is given  by

\be
C_{ij}^U
=C_{ij}^\prime \cos|y| +{\rm i}(C^\prime\gamma^\apr)_{ij}n^\apr\sin |y|
\  . \ee
This matrix is related to the standard charge conjugation matrix
$C_{ij}^\prime$
(which has  the properties: 
$C'^T=-C'$, $(C'\gamma^\apr)^T=-C'\gamma^\apr$, $C'^\dagger C'=1$)
 via a $y$-dependent 
unitarity transformation 
$C_{ij}^U = U^k{}_i C_{kl}^\prime U^l{}_j$   
(the explicit form of the matrix $U(y)$ is
given in  eq. (5.21) in  \ci{MT3}).

To transform  the superstring Lagrangian
 from the form  
(\ref{actkin3}),(\ref{actwz3}) into the one in 
(\ref{actkin6}),(\ref{actwz6})
we define the six-dimensional unit vector $u^M$ 

\be
u^6= \cos|y|\,,
\qquad \ \ \ 
u^\apr =\sin|y|n^\apr \ , 
\ee
and use (\ref{Cg}), (\ref{b2})--(\ref{b5})
which imply the following  relations

\begin{equation}\label{2rel}
G_{\sca\scb} (\eta V^\sca \eta) ( \eta V^\scb\eta)
=(\eta R^M\eta )^2\ , \ \ \ 
\qquad
G_{\sca\scb} (\eta V^\sca \eta) dy^\scb = 
(\eta R^M \eta) du^M \ , 
\end{equation}
where the $SU(4)$ rotation matrix $R^M$ is given by

\be
(R^M)^i{}_j =-\frac{1}{2}(\rho^{MN})^i{}_j u^N\,.
\ee
Note that $(R^M)^i{}_j\PP_M$ satisfies the 
commutation relations of the $su(4)$
algebra, i.e. plays the role of the angular
momentum operator $l^i_{\ j}$.

To relate the  WZ parts of the Lagrangians  (\ref{actwz3}) 
and (\ref{actwz6})
we use of the representation for the 
$SO(5)$  Dirac  and
$C^\prime$ matrices given in (\ref{Cg}) 
and thus find that  \ci{MT3}

\be
C_{ij}^U = \rho_{ij}^M u^M 
\ee
which implies the  desired transformation of the WZ parts
of the actions. 

It is often useful to replace the  unit six
dimensional vector $u^M$  by the  selfdual 
$SU(4)$  tensor $y_{ij}$  (or its inverse  $y^{ij}$) defined by 

\be
y_{ij} \equiv \rho_{ij}^M u^M\,,\ \ \ 
\qquad
y^{ij} \equiv (\rho^M)^{ij}u^M\ , \ \ \ \  
y^{il}y_{lj}=\delta^i_j\,. 
\ee
%which satisfy \be y^{il}y_{lj}=\delta^i_j\,. \ee
The relations \rf{iii} for  the $\rho^M$ matrices  imply that 

\be
y_{ij} = \frac{1}{2}\epsilon_{ijkl}y^{kl}\,,
\qquad
y_{ij}^* = - y^{ij}\,,
\qquad
y_{ij} = -y_{ji}\,.
\ee

%%%%%%%%%%%%%%%%%%%%%%%%%%%%%%%%%%%%%%%%%%%%%%%%%%%%%%%%%
\appendix{Manifestly $SU(4)$
 invariant  forms  of  phase space  \adss  Lagrangian}
%%%%%%%%%%%%%%%%%%%%%%%%%%%%%%%%%%%%%%%%

The  \lc  gauge superstring  action of \ci{MT3}
can be written in several equivalent forms
corresponding to different parametrizations of \adss space.
The two  forms corresponding to ``5+5" parametrizations 
 \rf{actkin3},\rf{actwz3} and 
\rf{actkin6},\rf{actwz6}
were discussed  in  Section 5.
Here we shall  review the  Lagrangians \ci{MT3}
for the two standard ``4+6" coordinate choices 
 and present  their phase space counterparts.

Choosing the 10 Cartesian coordinates $(x^a, Y^M)$ such that the metric
of $AdS_5\times S^5$  takes the form
\be
ds^2 = Y^2dx^a dx^a + { Y^{-2}} dY^M dY^M  \ , \ \ \ \ \  \ \ \ 
Y^2 = |Y|^2 = Y^M Y^M \ , 
\ee
one can  transform the bosonic coordinates
in   the ``intermediate" form of the superstring Lagrangian 
(\ref{actkin3}),(\ref{actwz3})  to obtain 
its   ``4+6"  form  \ci{MT3} 
$$ {\cal L} = {\cal L}_{kin} + {\cal L}_{WZ} \ ,  $$
$$
{\cal L}_{kin}
=
-\sqrt{g}g^{\vm\vn}\Bigl[
Y^2(\partial_\vm x^+ \partial_\vn x ^-
+ \partial_\vm x\partial_\vn\bar{x})
+\frac{1}{2} Y^{-2} D_\vm Y^M D_\vn Y^M\Bigr]
$$
\begin{equation}
- \ \frac{{\rm i}}{2} \sqrt{g}g^{\vm\vn}
Y^2\partial_\vm x^+
\Bigl[\theta^i\partial_\vn \theta_i
+\theta_i\partial_\vn \theta^i
+\eta^i\partial_\vn \eta_i
+\eta_i\partial_\vn \eta^i 
+{\rm i}  Y^2\partial_\vn x^+(\eta^2)^2\Bigr]\ ,
\label{actkin4n}
\end{equation}
\begin{equation}\label{actwz4n}
{\cal L}_{WZ}
=\epsilon^{\vm\vn}
|Y|\partial_\vm x^+ \eta^i \rho_{ij}^M Y^M
(\partial_\vn\theta^j-{\rm i}\sqrt{2}|Y| \eta^j
\partial_\vn x)+h.c. \ . 
\end{equation}
Here
\be
DY^M = dY^M -2 {\rm i}\eta_i (R^M)^i{}_j\eta^j  Y^2 dx^+\, ,
\qquad\ \  R^M  = - \ha \rho^{MN} Y^N \ , 
\ee
and the  matrices $\rho^M, \ \rho^{MN}$ were defined 
in \rf{usgam},\rf{rrr}.

It is easy to see that this Lagrangian can be represented in the 
same form as 
(\ref{lagdec}). Applying the final result for the phase 
space Lagrangian given in
(\ref{genph1})--(\ref{genph3})
we get
\begin{eqnarray}
{\cal L} 
&=& \PP_\al\dot{x}_\al 
+ \PP_M \dot{Y}^M
+\frac{{\rm i}}{2}p^+(\theta^i \dot{\theta}_i
+\eta^i\dot{\eta}_i+\theta_i \dot{\theta}^i+\eta_i\dot{\eta}^i)
\nonumber\\
&-&\frac{1}{2p^+}\Bigl[\PP_\al^2 +|Y|^4\PP_M\PP_M
+|Y|^4\x'_\al^2+ \Y'^M\Y'^M
+Y^2(p^{+2}(\eta^2)^2 + 4{\rm i}p^+\eta R^M\eta  \PP_M)\Bigr]
\nonumber\\
&+&\Bigl[\ |Y|\eta^i \rho_{ij}^M Y^M
(\th'^j - {\rm i}\sqrt{2}|Y| \eta^j\x')+h.c.\Bigr]
\nonumber\\
&-&
\frac{h^{01}}{p^+}\Bigl[p^+\x'^-  + \PP_\al\x'_\al 
+ \PP_M \Y'^M
+\frac{{\rm i}}{2}p^+(
\theta^i\th'_i+\eta^i\et'_i
+\theta_i\th'^i+\eta_i\et'^i)\Bigr]  \ . 
\la{phase4}
\end{eqnarray}
%\bigskip
%\begin{center}
%{\bf Phase action in conformal flat metric}
%\end{center}
The closely related choice of  4+6 coordinates 
is the one  which  makes explicit  the 
hidden spacetime  conformal symmetry of $AdS_5\times S^5$
geometry, i.e. the one in which the  metric 
takes the  conformally flat form

\be\label{cfc}
ds^2 =Z^{-2}(dx^a dx^a + dZ^MdZ^M) \ , \ \ \ \ \ \ \ \ \ 
 Z^M = {Y^M \ov Y^2} \ . 
\ee
%These coordinates manifest hidden spacetime 
%conformal symmetries of $AdS_5\times S^5$
%geometry. Though it is not clear the role these symmetries play in string theory
%it is helpful for future studies to discuss the Lagrangian in terms of these
%coordinates.
The superstring Lagrangian expressed in terms of 
these ``conformally   flat"   coordinates 
is readily obtained from \rf{actkin4n},\rf{actwz4n}
 \ci{MT3}

$$
{\cal L}_{kin}
=
-\sqrt{g}g^{\vm\vn}Z^{-2}\Bigl[
\partial_\vm x^+ \partial_\vn x ^-
+ \partial_\vm x\partial_\vn\bar{x}
+\frac{1}{2}D_\vm Z^M D_\vn Z^M\Bigr]
$$
\begin{equation}
- \ \frac{{\rm i}}{2} \sqrt{g}g^{\vm\vn}
Z^{-2}\partial_\vm x^+
\Bigl[\theta^i\partial_\vn \theta_i
+\theta_i\partial_\vn \theta^i
+\eta^i\partial_\vn \eta_i
+\eta_i\partial_\vn \eta^i 
+{\rm i}  Z^{-2}\partial_\vn x^+(\eta^2)^2\Bigr]\ ,
\label{actkin5n}
\end{equation}
\begin{equation}\label{actwz5n}
{\cal L}_{WZ}
=\epsilon^{\vm\vn}
|Z|^{-3}\partial_\vm x^+ \eta^i \rho_{ij}^M Z^M
(\partial_\vn\theta^j-{\rm i}\sqrt{2}|Z|^{-1} \eta^j
\partial_\vn x)+h.c.\ , 
\end{equation}
where
\be 
DZ^M = dZ^M -2 {\rm i}\eta_i (R^{M})^i{}_j\eta^j  Z^{-2} dx^+\, ,
\qquad\ \  R^M  = - \ha \rho^{MN} Z^N \ .
\ee 
%$$DZ^M = dZ^M -2{\rm i}\eta_i (R^M)^i{}_j\eta^j Z^{-2} dx^+$$
This Lagrangian has again  the same  form as in
(\ref{lagdec})--(\ref{l3}), and its phase space counterpart is
thus found from  (\ref{ll1})--(\ref{ll3})

\begin{eqnarray}
{\cal L} 
&=& \PP_\al\dot{x}_\al 
+ \PP_M \dot{Z}^M
+\frac{{\rm i}}{2}p^+(\theta^i \dot{\theta}_i
+\eta^i\dot{\eta}_i+\theta_i \dot{\theta}^i+\eta_i\dot{\eta}^i)
\nonumber\\
&-&\frac{1}{2p^+}\Bigl[\PP_\al^2 +\PP_M\PP_M
+Z^{-4}(\x'_\al^2+ \Z'^M\Z'^M)
+Z^{-2}(p^{+2}(\eta^2)^2 + 4{\rm i}p^+\eta R^M\eta \PP_M)\Bigr]
\nonumber\\
&+&\Bigl[\ |Z|^{-3}\eta^i \rho_{ij}^M Z^M
(\th'^j - {\rm i}\sqrt{2}|Z|^{-1} \eta^j\x')+h.c.\Bigr]
\nonumber\\
&-&
\frac{h^{01}}{p^+}\Bigl[p^+\x'^-  + \PP_\al\x'_\al 
+ \PP_M \Z'^M
+\frac{{\rm i}}{2}p^+(
\theta^i\th'_i+\eta^i\et'_i
+\theta_i\th'^i+\eta_i\et'^i)\Bigr] \ . 
\end{eqnarray}
An interesting feature of this Lagrangian is that the squares
 of the $AdS_5$ and $S^5$  momenta  enter exactly 
  as in  the flat space case.

%Note that aslo here the square of
%derivatives of $AdS$ transverse coordinates  (i.e. $\x'_al^2$) 
%and the one of $S^5$ part (i.e. $\Z'_M^2$) enter with the same measure.

%%%%%%%%%%%%%%%%%%%%%%%%%%%%%%%%%%%%%%%%%%
\appendix{Conformal supercharges}
%%%%%%%%%%%%%%%%%%%%%%%%%%%%%%%%%%%%

In order to find the conformal supercharges supplementing 
the generators given in Section 6,  one needs to know 
the contribution  of the 
compensating $\k$-symmetry transformation  given in \ci{MT1}. 
Finding these
compensating transformations directly is 
rather  complicated.
% and not very illuminating procedure.
 Here we suggest an indirect  method based on exploiting the  known
equations of motion given in (\ref{eqmot1})--(\ref{eqmot11}) and
the commutation relations of the $psu(2,2|4)$ superalgebra. 
Let us demonstrate  this 
procedure for the case of the  conformal supercharges $S^{+i}$.

In general, the  charges depend on the 
 time variables $x^+=\tau$  explicitly and also 
through the  dynamical variables, i.e. $G=G(\tau, {\cal X}(\tau))$. 
Because our
equations of motion (\ref{eqmot1})--(\ref{eqmot11}) have
a  Hamiltonian form
(\ref{hamfor}) we can rewrite the  conservation law
$
\frac{dG}{dx^+}= 0
$
as follows (as in \rf{hamfor}--\rf{qcom6}
here $[\ , \ ]$ stand for the classical Poisson
bracket)

\be
\frac{\partial G}{\partial x^+}+[G,P^-]=0 \ . 
\ee
This determines the explicit  dependence on $x^+$, 
namely, 
\be\label{mandep}
G_{x^+}=G_{x^+=0}
+x^+ [P^-,G_{x^+=0}]
+\frac{x^{+2}}{2}[P^-,[P^-,G_{x^+=0}]] \ , 
\ee
where we used the  notation

\be\label{g0}
G|_{x^+ =0}
\equiv 
G(0,{\cal X}(\tau)) \ , 
\ee
and took into account that the
 expansion in $x^+$ terminates at   $(x^{+})^2$ order
because the $psu(2,2|4)$ superalgebra
does not have generators  with  absolute  value of the 
$J^{+-}$ charge higher than 1
(so that triple and higher commutators of $P^-$
with any generator vanish).

For example,  eq. (\ref{mandep})   implies that 
the  charges $J^{+x}$, $J^{+-}$, $D$
depend on $x^+$    as follows

\begin{equation}\label{adsjpcom}
J^{+ x}=  J^{+ x}|_{x^+=0} + x^+ P\,,
\quad\ \ \ 
J^{+\bar x}=  J^{+\bar{x}}|_{x^+=0} + x^+ \bar{P}\ , 
\end{equation}
\be
J^{+-}=J^{+-}|_{x^+=0} + x^+ P^-,
\qquad\ \ 
D=D|_{x^+=0}+ x^+ P^-.
\end{equation}
These expressions match  the ones given in (\ref{kin2}),(\ref{diag3}),(\ref{diag4}).
We also learn  from (\ref{mandep}) 
and the  commutation relations of $psu(2,2|4)$ 
that the dynamical
generators and $J^i{}_j$, $J^{x\bar{x}}$
do not explicitly depend  on $x^+$ (cf. (\ref{diag1}),(\ref{diag2})).

Let us now  turn to the conformal supercharges. From (\ref{mandep}) and
commutation relations of $psu(2,2|4)$  we find

\be\label{app0}
S^+_i=S_i^+|_{x^+=0} -{\rm i}x^+ Q^-_i\,,
\qquad
S^{+i}=S^{+i}|_{x^+=0} + {\rm i}x^+ Q^{-i}\,, 
\ee

\be
Q^-_i|_{x^+=0} = Q^-_i\,,
\qquad
Q^{-i}|_{x^+=0} = Q^{-i}\,,
\ee
and  thus

\be\label{spq}
[S^+_i|_{x^+=0},P^-]={\rm i}Q_i^-\,,
\qquad
[S^{+i}|_{x^+=0},P^-]=-{\rm i}Q^{-i}\,.
\ee
As a result,  in order to determine  the conformal supercharges
 $S^+_i$ we have to
find only their $S^+_i|_{x^+=0}$ part. 
In the flat space case 
 this ${x^+=0}$ part of the kinematic generators is
obtainable simply from  the particle charges by replacing 
the particle coordinates
by the ones of the  string. 
The same can be done 
in  the case of
$AdS$ space. 
Here  we start with the 
superparticle expressions for the conformal  supercharges $S^+_i$ given in 
\ci{met3} and
replace the superparticle coordinates by the superstring ones

\be\label{app1}
{\cal S}_i^{+ 0}|_{x^+=0}
=\frac{1}{\sqrt{2}}e^{-\phi} p^+\eta_i +{\rm i}p^+\theta_i x\,.
\ee
Using the analog of   (\ref{app0}) for the currents 

\be\label{app2}
{\cal S}^{+0}_i={\cal S}_i^{+0}|_{x^+=0} -{\rm i}x^+ {\cal Q}^{-0}_i\,,
\ee
and  the conservation law for  ${\cal S}_i^{+\mu}$  and 
${\cal Q}_i^{+\mu}$  one finds 
the following expressions for 
 the conformal supercurrent ${\cal S}_i^{+1}$ 
and the Poincar\'e supercurrent ${\cal Q}^{-\mu}_i$

\be
{\cal S}_i^{+ 1}
=
e^{2\phi} y_{ij}\eta^j x -{\rm i}x^+{\cal Q}_i^{-1}\,,
\qquad
{\cal S}^{+i 1}
= -e^{2\phi} y^{ij} \eta_j\bar{x} + {\rm i}x^+{\cal Q}^{-i1}\,, 
\ee
\be
{\cal Q}_i^{-0} = \PP \theta_i 
+\frac{e^\phi}{\sqrt{2}}
\Bigl({\rm i}\eta_i\Pi
-p^+\eta^2\eta_i + 2 (\eta l)_i
+y_{ij}(\th'^j -{\rm i}\sqrt{2}e^\phi\eta^j\x')\Bigr)\ , 
\ee
\be
{\cal Q}_i^{-1}
=e^{3\phi}\theta_i (-\frac{e^{\phi}}{p^+}\x'
+{\rm i}\sqrt{2}\eta_k y^{kl} \eta_l)
+\frac{\rm ie^{2\phi}}{p^+}y_{ij}\eta^j\PP
+\frac{\rm ie^{2\phi}}{\sqrt{2}p^+}\partial_\sigma (e^\phi\eta_i)
+2\sqrt{2}e^\phi \eta_j (l^j{}_i)^1\,,
\ee
where
\be
(l^i{}_j)^1 \equiv
 -\frac{\rm i}{2p^+}e^{2\phi}(\rho^{MN})^i{}_j u^M \u'^N \ . 
\ee
These relations determine the  supercharges.
Finally,  one can check  that these charges  satisfy the  
commutation relations of the $psu(2,2|4)$ superalgebra. 
The corresponding commutation relations
are obtainable from the
ones given in Section  3 of  \ci{MT3} by making the following 
substitutions there:
$J^i{}_j \rightarrow -{\rm i}J^i{}_j$, 
$P^a\rightarrow -P^a$,
$K^a\rightarrow - K^a$, $S\rightarrow -S$.

%%%%%%%%%%%%%%%%%%%%%%%%%%%%%%%

\end{document}